\documentclass[aps,prd,twocolumn,superscriptaddress]{revtex4}
\usepackage{epsfig,epsf}
\usepackage{amsmath}
\usepackage{amsthm}
\usepackage{amsfonts}
\usepackage{amssymb}
\usepackage{dsfont}
\usepackage{multirow}
\usepackage{appendix}
\usepackage{slashed}
\usepackage[active]{srcltx}
\usepackage{psfrag}

\begin{document}

\title{Nature of the vector resonance $Y(2175)$}
\date{\today}
\author{S.~S.~Agaev}
\affiliation{Institute for Physical Problems, Baku State University, Az--1148 Baku,
Azerbaijan}
\author{K.~Azizi}
\affiliation{Department of Physics, University of Tehran, North Karegar Avenue, Tehran
14395-547, Iran}
\affiliation{Department of Physics, Do\v{g}u\c{s} University, Acibadem-Kadik\"{o}y, 34722
Istanbul, Turkey}
\author{H.~Sundu}
\affiliation{Department of Physics, Kocaeli University, 41380 Izmit, Turkey}

\begin{abstract}
Spectroscopic parameters and decay channels of the vector resonance $Y(2175)$
are studied by considering it as a diquark-antidiquark state with the quark
content $su\overline{s}\overline{u}$. The mass and coupling of the
tetraquark $Y(2175)$ are calculated using the QCD two-point sum rules by
taking into account various quark, gluon and mixed condensates up to
dimension 15. Partial widths of its strong decays to $\phi f_{0}(980)$, $
\phi \eta$, and $\phi \eta^{\prime}$ are computed as well. To this end, we
explore the vertices $Y\phi f_{0}(980)$, $Y\phi \eta$, and $Y\phi
\eta^{\prime}$, and calculate the corresponding strong couplings by means of
the QCD light-cone sum rule method. The coupling $G_{Y\phi f}$ of the vertex
$Y\phi f_{0}(980)$ is found using the full version of this method, and by
treating the scalar meson $f_{0}(980)$ as a diquark-antidiquark tetraquark
state. The couplings $g_{Y\phi \eta }$ and $g_{Y\phi \eta^{\prime}}$,
however, are calculated by applying the soft-meson approximation to the
light-cone sum rule method. Prediction for the mass of the resonance $m_{Y}
=\left( 2173\pm 85\right)~\mathrm{MeV}$ is in excellent agreement with the
data of the BaBar Collaboration \cite{Aubert:2006bu}, and within errors of
calculations is compatible with the result reported by BESIII \cite
{Ablikim:2014pfc}. The full width $\Gamma _{\mathrm{full}}=(91.1\pm 20.5)~
\mathrm{MeV}$ of the $Y(2175)$ saturated by its three strong decay channels
is in a reasonable agreement with existing experimental data.
\end{abstract}

\maketitle

%%%%%%%%%%%%%%%%%%%%%%%%%%%%%%%%%%%%%%%%%%%%%%%%%%%%%%%%%%%%%%

\section{Introduction}

\label{sec:Int}
%%%%%%%%%%%%%%%%%%%%%%%%%%%%%%%%%%%%%%%%%%%%%%%%%%%%%%%%%%%%%%
The resonances $\{Y\}$ with the quantum numbers $J^{\mathrm{PC}}=1^{--}$
constitute two families of particles, interpretation of which is one of
interesting and yet unsettled problems of the high energy physics. Members
of the first family populate the mass region $m=4.2-4.7~\mathrm{GeV}$, and
were observed by different collaborations. These resonances reside very
close to each other, and are more numerous than vector charmonia $\overline{c%
}c$ from this mass range. Hence, at least some of these resonances have
different quark-gluon structure, and are presumably states built of four
valence quarks. Besides a suggestion about the tetraquark nature of heavy $%
\{Y\}$ states, there are various alternative models to account for their
parameters and decay channels.

Another family of the $\{Y\}$ resonances occupies the light segment of meson
spectroscopy and incorporates the famous "old" state $Y(2175)$, and new ones
$X(2239)$ and $X(2100)$ seen recently. The structure $Y(2175)$ was
discovered by the BaBar Collaboration in the initial-state radiation process
$e^{+}e^{-}\rightarrow \gamma _{\mathrm{ISR}}\phi f_{0}(980)$ as a resonance
in the $\phi f_{0}(980)$ invariant mass spectrum \cite{Aubert:2006bu}. The
mass and width of this resonance measured by BaBar amount to $m=(2175\pm
10\pm 15)~\mathrm{MeV}$ and $\Gamma =(58\pm 16\pm 20)~\mathrm{MeV}$,
respectively. The same structure was seen also by the BESIII Collaboration
in the exclusive decay $J/\psi \rightarrow \eta \phi \pi ^{+}\pi ^{-}$ \cite%
{Ablikim:2014pfc}. The spectroscopic parameters of the $Y(2175)$ extracted
in this experiment differ from original results and are $m=(2200\pm 6\pm 5)~%
\mathrm{MeV}$ and $\Gamma =(104\pm 15\pm 15)~\mathrm{MeV}$. Recently,
anomalously high cross section at $\sqrt{s}=2232~\mathrm{MeV}$ was observed
by the BESIII Collaboration in the channel $e^{+}e^{-}\rightarrow \phi
K^{+}K^{-}$, which may be explained by interference of different resonances
\cite{Ablikim:2019tpp}: more data are necessary to decide whether $Y(2175)$
contributes to enhancement of this cross section or not. Because the $%
Y(2175) $ was seen by BaBar and confirmed by the BESII, BESIII, and Belle
collaborations \cite{Ablikim:2007ab,Shen:2009zze,Ablikim:2014pfc}, its
existence is not in doubt, but an uncertain situation with the mass and full
width of \ this resonance requires further experimental and theoretical
studies.

Other resonances that may be considered as candidates to light exotic vector
mesons were discovered by the BESIII Collaboration. The first of them, i.e.,
$X(2239)$ was fixed in the process $e^{+}e^{-}\rightarrow K^{+}K^{-}$ as a
resonant structure in the cross section shape line \cite{Ablikim:2018iyx}.
The second resonance $X(2100)$ was seen in the $\phi \eta ^{\prime }$ mass
spectrum in the process $J/\psi \rightarrow \phi \eta \eta ^{\prime }$ \cite%
{Ablikim:2018xuz}. The quantum numbers of $X(2239)$ were determined
unambiguously, whereas a situation with $X(2100)$ remains unclear. Indeed,
because of a scarcity of experimental information the collaboration could
not clearly distinguish two $1^{+}$ and $1^{-}$ assumptions for the
spin-parity $J^{P}$ of the resonance $X(2100)$. Hence, BESIII extracted its
mass and full width using both of these options. Obtained results differ
from each other and depend on assumption about the parity of the state $%
X(2100)$.

Theoretical interpretations of the light vector resonances comprise all
available models and approaches of the high energy physics. Thus, the $%
Y(2175)$ was considered as $2{}^{3}D_{1}$ excitation of the ordinary $%
\overline{s}s$ meson \cite{Ding:2007pc,Wang:2012wa}. It was explained also
as a dynamically generated state in the $\phi K\overline{K}$ system \cite%
{MartinezTorres:2008gy}, or as a resonance appeared due to self-interaction
between $\phi $ and $f_{0}(980)$ mesons \cite{AlvarezRuso:2009xn}. A hybrid
meson with structure $\overline{s}sg$ \cite{Ho:2019org} and a
baryon-antibaryon $qqs\overline{q}\overline{q}\overline{s}$ state which
couples strongly to the $\Lambda \overline{\Lambda }$ channel are among
alternative models of the $Y(2175)$ resonance. There were attempts to
interpret $Y(2175)$ as a vector tetraquark with $s\overline{s}s\overline{s}$
or $ss\overline{s}\overline{s}$ contents \cite%
{Wang:2006ri,Chen:2008ej,Chen:2018kuu} (see Ref.\ \cite{Ablikim:2018iyx} for
other models). The resonance $X(2100)$ was examined in the framework of the
QCD sum rule method in Refs.\ \cite{Cui:2019roq,Wang:2019nln}.

Recently, we explored the light resonances $X(2100)$ and $X(2239)$ as the
axial-vector and vector $ss\overline{s}\overline{s}$ tetraquarks \cite%
{Azizi:2019ecm}, respectively. Besides spectroscopic parameters we also
investigated the strong decays $X(2100)\rightarrow \phi \eta ^{\prime }$ and
$X(2100)\rightarrow \phi \eta $, and calculated their partial widths.
Predictions obtained for the mass and width of the axial-vector state
allowed us to identify it with the resonance $X(2100)$, because our
theoretical predictions are very close to its parameters measured by the
BESIII Collaboration. We classified $X(2239)$ as the vector tetraquark $ss%
\overline{s}\overline{s}$ and found a reasonable agreement between
theoretical and experimental results.

In the present work, we continue our investigations of the light vector
resonances and concentrate on features of the state $Y(2175)$ (hereafter, $Y$%
). Our treatment of this state differs from existing analyses. Thus, we
consider it as a vector tetraquark with content $su\overline{s}\overline{u}$
rather than as a state $ss\overline{s}\overline{s}$. The traditional
assumption about the quark content of the $Y$ is inspired by the fact, that
it was discovered in $\phi f_{0}(980)$ invariant mass distribution. Because
in the standard model of mesons one treats the $\phi $ and $f_{0}(980)$ as
vector and scalar particles with the same $\overline{s}s$ structure, then it
is natural to assume that $Y$ is built of four valence $s$ quarks.

But the conventional quark-antiquark model of mesons in the case of light
scalar nonets meets with evident difficulties. In fact, the nonet of scalar
mesons in the $\overline{q}q$ model may be realized as $1{}^{3}P_{0}$
states. In accordance with various computations, masses of the scalars $%
1{}^{3}P_{0}$ are higher than $1~\mathrm{GeV}$. They were identified with
the isoscalar mesons $f_{0}(1370)$ and $f_{0}(1710)$, the isovector $%
a_{0}(1450)$ or isospinor $K_{0}^{\ast }(1430)$ states, i.e., with scalars
from the second light nonet. But masses of the mesons from the first nonet
are lower than $1\ \mathrm{GeV}$, and they cannot be included into this
scheme. Therefore, to explain experimental information on their masses, and
an unusual mass hierarchy inside of the nonet ~Jaffe made a suggestion on a
four-quark nature of these particles \cite{Jaffe:1976ig}.

An updated model of the light scalar nonets is based on assumption about a
diquark-antidiquark structure of these particles, which appear as mixtures
of spin-$0$ diquarks from ($\overline{\mathbf{3}}_{c},\overline{\mathbf{3}}%
_{f}$) representation with spin-$1$ diquarks from ($\mathbf{6}_{c},\
\overline{\mathbf{3}}_{f})$ representation of the color-flavor group \cite%
{Kim:2017yvd}. In Refs.\ \cite{Agaev:2017cfz,Agaev:2018sco} we investigated
the scalar mesons $f_{0}(500)$ and $f_{0}(980)$ as admixtures of the $%
SU_{f}(3)$ basic light $\mathbf{L}=[ud][\overline{u}\overline{d}]$ and heavy
$\mathbf{H}=([su][\overline{s}\overline{u}]+[ds][\overline{d}s])/\sqrt{2}$
tetraquark states, and calculated their spectroscopic parameters and full
widths. Obtained predictions agree with existing experimental data,
therefore we consider the $f_{0}(980)$ as the exotic four-quark meson. Once
we accept this model, a treatment of the $Y$ as a vector tetraquark $Y=[su][%
\overline{s}\overline{u}]$ becomes quite reasonable.

We calculate the spectroscopic parameters of the vector tetraquark $Y=[su][%
\overline{s}\overline{u}]$ and explore some of its decay channels. The mass
and coupling of the $Y$ are evaluated using the QCD two-point sum rule
method \cite{Shifman:1978bx,Shifman:1978by}. We investigate the strong
decays $Y\rightarrow $ $\phi f_{0}(980)$,$\ Y\rightarrow $ $\phi \eta $, and
$Y\rightarrow $ $\phi \eta ^{\prime }$, and find their partial widths. To
this end, we use the QCD light-cone sum rule (LCSR) method \cite%
{Balitsky:1989ry}, and calculate the couplings $G_{Y\phi f}$, $g_{Y\phi \eta
}$, and $g_{Y\phi \eta ^{\prime }}$ corresponding to the strong vertices $%
Y\phi f_{0}(980)$, $Y\phi \eta $, and $Y\phi \eta ^{\prime }$, respectively.
The coupling $G_{Y\phi f}$ is computed by employing the full version of the
LCSR method, whereas in the case of $g_{Y\phi \eta }$, and $g_{Y\phi \eta
^{\prime }}$ this method is supplemented by a technique of the soft-meson
approximation \cite{Belyaev:1994zk,Ioffe:1983ju,Agaev:2016dev}. Because the
light component of $f_{0}(980)$ is irrelevant for analysis of the decay $%
Y\rightarrow \phi f_{0}(980)$, we treat $f_{0}(980)$ as a pure $\mathbf{H}$
state.

This article is organized as the following way: In Sec.\ \ref{sec:Mass} we
calculate the mass and coupling of the tetraquark $Y$. The strong decays of
this state are considered in sections \ref{sec:Decay1} and \ref{sec:Decay2}.
In Sec.\ \ref{sec:Decay1} we analyze the process $Y\rightarrow \phi
f_{0}(980)$ using the LCSR method and find partial decay width of this
channel. The partial widths of the decay modes $Y\rightarrow $ $\phi \eta $%
,\ and $Y\rightarrow $ $\phi \eta ^{\prime }$ are calculated in the next
section \ref{sec:Decay2}. In Sec.\ \ref{sec:Conclusions} we analyze the
obtained results, and give our conclusions.

%%%%%%%%%%%%%%%%%%%%%%%%%%%%%%%%%%%%%%%%%%%%%%%%%%%%%%%%%%%%%%%%%%%%

\section{Spectroscopic parameters of the tetraquark $Y$: the mass $m_{Y}$
and current coupling $f_{Y}$}

\label{sec:Mass}
%%%%%%%%%%%%%%%%%%%%%%%%%%%%%%%%%%%%%%%%%%%%%%%%%%%%%%%%%%%%%%%%

To evaluate the mass $m_{Y}$ and coupling\ $f_{Y}$ of the vector tetraquark $%
Y$, we use the QCD two-point sum rule method and start our calculations from
analysis of the correlation function%
\begin{equation}
\Pi _{\mu \nu }(p)=i\int d^{4}xe^{ipx}\langle 0|\mathcal{T}\{J_{\mu
}^{Y}(x)J_{\nu }^{Y\dag }(0)\}|0\rangle ,  \label{eq:CF1}
\end{equation}%
where $J_{\mu }^{Y}(x)$ is the interpolating current for the $Y$ state.

The current for a tetraquark with $J^{\mathrm{P}}=1^{-}$ can be built of a
scalar diquark and vector antidiquark or/and a vector diquark and scalar
antidiquark. There are several options to construct alternative currents
with required spin-parities, but because a scalar diquark (antidiquark) is a
most stable two-quark state \cite{Jaffe:2004ph}, for $J_{\mu }^{Y}$ we use
the structure
\begin{equation}
C\gamma _{5}\otimes \gamma _{\mu }\gamma _{5}C-C\gamma _{\mu }\gamma
_{5}\otimes \gamma _{5}C.  \label{eq:Structure}
\end{equation}%
This current consists of two components, and each of them describes a vector
tetraquark. The whole structure corresponds to a vector tetraquark with
definite charge-conjugation parity $J^{\mathrm{PC}}=1^{--}$. Indeed, the
charge-conjugation transforms diquarks to antidiquarks and vice versa,
therefore the minus sign between two components in Eq.\ (\ref{eq:Structure})
generates the current with $C=-1$.

The last question to be solved is a color structure of constituent diquarks
and antidiquarks. Thus, to get the color-singlet current $J_{\mu }^{Y}$ they
should have the same color structures and be either in color triplet $[\overline{%
\mathbf{3}}_{c}]\otimes \lbrack \mathbf{3}_{c}]$ or sextet $[\mathbf{6}%
_{c}]\otimes \lbrack \overline{\mathbf{6}}_{c}]$ configurations. The current
of the type (\ref{eq:Structure}) and built of color-sextet
diquark-antidiquark has the following form \cite{Chen:2010ze}%
\begin{eqnarray}
J_{1\mu } &=&u_{a}^{T}C\gamma _{5}s_{b}[\overline{u}_{a}\gamma _{\mu }\gamma
_{5}C\overline{s}_{b}^{T}+\overline{u}_{b}\gamma _{\mu }\gamma _{5}C%
\overline{s}_{a}^{T}]  \notag \\
&&-u_{a}^{T}C\gamma _{\mu }\gamma _{5}s_{b}[\overline{u}_{a}\gamma _{5}C%
\overline{s}_{b}^{T}+\overline{u}_{b}\gamma _{5}C\overline{s}_{a}^{T}].
\label{eq:CurrSext}
\end{eqnarray}%
The triplet current (\ref{eq:Structure}) is given by the expression%
\begin{eqnarray}
J_{3\mu } &=&u_{a}^{T}C\gamma _{5}s_{b}[\overline{u}_{a}\gamma _{\mu }\gamma
_{5}C\overline{s}_{b}^{T}-\overline{u}_{b}\gamma _{\mu }\gamma _{5}C%
\overline{s}_{a}^{T}]  \notag \\
&&-u_{a}^{T}C\gamma _{\mu }\gamma _{5}s_{b}[\overline{u}_{a}\gamma _{5}C%
\overline{s}_{b}^{T}-\overline{u}_{b}\gamma _{5}C\overline{s}_{a}^{T}].
\label{eq:CurrTrip}
\end{eqnarray}%
In Eqs.\ (\ref{eq:CurrSext}) and (\ref{eq:CurrTrip}) $a$ and $b$ are color
indices, and $C$ is the charge-conjugation matrix.

The $J_{1\mu }$ and $J_{3\mu }$ are color-singlet currents composed of
color-sextet and -triplet diquark-antidiquark pairs, respectively. To see
this, let us consider in a detailed form  $J_{1\mu }$. The color-sextet,
i.e., color-symmetric $a\leftrightarrow b$ nature of the antidiquark fields
in Eq.\ (\ref{eq:CurrSext}) is evident. The first component of $J_{1\mu }$,
for example, in the explicit color-singlet form is
\begin{equation}
(u_{a}^{T}C\gamma _{5}s_{b}+u_{b}^{T}C\gamma _{5}s_{a})[\overline{u}%
_{a}\gamma _{\mu }\gamma _{5}C\overline{s}_{b}^{T}+\overline{u}_{b}\gamma
_{\mu }\gamma _{5}C\overline{s}_{a}^{T}],  \label{eq:ColS}
\end{equation}%
where both the diquark and antidiquark are symmetric in color indices. It is
not difficult to see, that diquarks $u_{a}^{T}C\gamma _{5}s_{b}$ and $%
u_{b}^{T}C\gamma _{5}s_{a}$ lead to identical results, hence it is enough in
$J_{1\mu }$ to keep one of them. The similar analysis is valid for the
second component of $J_{1\mu }$ as well. In the case of the current $J_{3\mu
}$, we see that the antidiquark fields in Eq. (\ref{eq:CurrTrip}) are
color-triplet or color-antisymmetric constructions. The color-triplet
diquark field, for example, in the first component of $J_{3\mu }$ is $%
(u_{a}^{T}C\gamma _{5}s_{b}-u_{b}^{T}C\gamma _{5}s_{a})$, and both $%
u_{a}^{T}C\gamma _{5}s_{b}$ and $-u_{b}^{T}C\gamma _{5}s_{a}$ give again the
same results. Therefore, we  use one of them in the current $J_{3\mu }$ and
get (\ref{eq:CurrTrip}).

An appropriate form of the current $J_{\mu }^{Y}$ that ensures stability and
convergence of the sum rules, which are actual in the case of light
tetraquarks \cite{Chen:2007xr}, is superposition of $J_{1\mu }$ and $J_{3\mu
}$. In the present work we use $J_{\mu }^{Y}=(J_{1\mu }+J_{3\mu })/2$, and
get%
\begin{eqnarray}
J_{\mu }^{Y}(x) &=&[u_{a}^{T}(x)C\gamma _{5}s_{b}(x)][\overline{u}%
_{a}(x)\gamma _{\mu }\gamma _{5}C\overline{s}_{b}^{T}(x)]  \notag \\
&&-[u_{a}^{T}(x)C\gamma _{\mu }\gamma _{5}s_{b}(x)][\overline{u}%
_{a}(x)\gamma _{5}C\overline{s}_{b}^{T}(x)].  \label{eq:Curr}
\end{eqnarray}%
The $J_{\mu }^{Y}(x)$ is a sum of two colorless terms, but belongs
neither to sextet nor to triplet representations of the color-group being
the admixture of such states $J_{1\mu }$ and $J_{3\mu }$.

To obtain sum rules for the mass and coupling of $Y$, we should express the
correlation function in terms of these spectral parameters, and also
calculate $\Pi _{\mu \nu }(p)$ using quark-gluon degrees of freedom. The
first expression forms the physical side of the sum rules $\Pi _{\mu \nu }^{%
\mathrm{Phys}}(p)$, whereas the second one constitutes their QCD side $\Pi
_{\mu \nu }^{\mathrm{OPE}}(p)$. In terms of the tetraquark's parameters the
correlation function has the following form

\begin{equation}
\Pi _{\mu \nu }^{\mathrm{Phys}}(p)=\frac{\langle 0|J_{\mu }^{Y}|Y(p)\rangle
\langle Y(p)|J_{\nu }^{Y\dagger }|0\rangle }{m_{Y}^{2}-p^{2}}+\dots
\label{eq:CF2}
\end{equation}%
Equation (\ref{eq:CF2}) is derived by saturating the correlation function
with a complete set of $J^{\mathrm{PC}}=1^{--}$ states and carrying out
integration in Eq.\ (\ref{eq:CF1}) over $x$. As usual, contributions arising
from higher resonances and continuum states are denoted above by dots.

The correlator $\Pi _{\mu \nu }^{\mathrm{Phys}}(p)$ can be further
simplified if one introduces the matrix element
\begin{equation}
\langle 0|J_{\mu }^{Y}|Y(p)\rangle =f_{Y}m_{Y}\epsilon _{\mu },
\label{eq:MElem1}
\end{equation}%
where $\epsilon _{\mu }$ is the polarization vector of the $Y$ state. Then
the correlation function $\Pi _{\mu \nu }^{\mathrm{Phys}}(p)$ takes the
simple form
\begin{equation}
\Pi _{\mu \nu }^{\mathrm{Phys}}(p)=\frac{m_{Y}^{2}f_{Y}^{2}}{m_{Y}^{2}-p^{2}}%
\left( -g_{\mu \nu }+\frac{p_{\mu }p_{\nu }}{m_{Y}^{2}}\right) +\ldots ,
\label{eq:CorF1}
\end{equation}%
and contains the Lorentz structure corresponding to the vector state.
Because a part of this structure proportional to $g_{\mu \nu }$ receives
contribution only from the vector states, we work with this term and
corresponding invariant amplitude $\Pi ^{\mathrm{Phys}}(p^{2})$.

The QCD side of the sum rules is given by the same correlation function $\Pi
_{\mu \nu }(p)$ but expressed in terms of the quark propagators.
Substituting the interpolating current into Eq.\ (\ref{eq:CF1}), and
contracting the quark fields, we get%
\begin{eqnarray}
&&\Pi _{\mu \nu }^{\mathrm{OPE}}(p)=i\int d^{4}xe^{ipx}\left\{ \mathrm{Tr}%
\left[ \gamma _{5}\widetilde{S}_{s}^{b^{\prime }b}(-x)\gamma _{5}\gamma
_{\nu }S_{u}^{a^{\prime }a}(-x)\right] \right.  \notag \\
&&\times \mathrm{Tr}\left[ S_{u}^{aa^{\prime }}(x)\gamma _{5}\widetilde{S}%
_{s}^{bb^{\prime }}(x)\gamma _{5}\gamma _{\mu }\right] +\mathrm{Tr}\left[
\gamma _{\mu }\gamma _{5}\widetilde{S}_{s}^{b^{\prime }b}(-x)\gamma
_{5}\right.  \notag \\
&&\left. \times S_{u}^{a^{\prime }a}(-x)\right] \mathrm{Tr}\left[
S_{u}^{aa^{\prime }}(x)\gamma _{\nu }\gamma _{5}\widetilde{S}%
_{s}^{bb^{\prime }}(x)\gamma _{5}\right] +\mathrm{Tr}\left[
S_{u}^{aa^{\prime }}(x)\gamma _{5}\right.  \notag \\
&&\left. \times \widetilde{S}_{s}^{bb^{\prime }}(x)\gamma _{5}\right]
\mathrm{Tr}\left[ \gamma _{\mu }\gamma _{5}\widetilde{S}_{s}^{b^{\prime
}b}(-x)\gamma _{5}\gamma _{\nu }S_{u}^{a^{\prime }a}(-x)\right]  \notag \\
&&+\mathrm{Tr}\left[ \gamma _{5}\widetilde{S}_{s}^{b^{\prime }b}(-x)\gamma
_{5}S_{u}^{a^{\prime }a}(-x)\right]  \notag \\
&&\left. \times \mathrm{Tr}\left[ S_{u}^{aa^{\prime }}(x)\gamma _{\nu
}\gamma _{5}\widetilde{S}_{s}^{bb^{\prime }}(x)\gamma _{5}\gamma _{\mu }%
\right] \right\} ,  \label{eq:CF3}
\end{eqnarray}%
where%
\begin{equation}
\widetilde{S}_{q}(x)=CS_{q}^{T}(x)C.  \label{eq:Prop}
\end{equation}%
In the formula above $S_{q}(x)$ is the light quark propagator, for which we
employ the expression
\begin{eqnarray}
&&S_{q}^{ab}(x)=i\frac{\slashed x}{2\pi ^{2}x^{4}}\delta _{ab}-\frac{m_{q}}{%
4\pi ^{2}x^{2}}\delta _{ab}-\frac{\langle \overline{q}q\rangle }{12}\left(
1-i\frac{m_{q}}{4}\slashed x\right) \delta _{ab}  \notag \\
&&-\frac{x^{2}}{192}\langle \overline{q}g_{s}\sigma Gq\rangle \left( 1-i%
\frac{m_{q}}{6}\slashed x\right) \delta _{ab}-\frac{\slashed xx^{2}g_{s}^{2}%
}{7776}\langle \overline{q}q\rangle ^{2}\delta _{ab}  \notag \\
&&-\frac{ig_{s}G_{ab}^{\mu \nu }}{32\pi ^{2}x^{2}}\left[ \slashed x\sigma
_{\mu \nu }+\sigma _{\mu \nu }\slashed x\right] -\frac{x^{4}\langle
\overline{q}q\rangle \langle g_{s}^{2}G^{2}\rangle }{27648}\delta _{ab}
\notag \\
&&+\frac{m_{q}g_{s}}{32\pi ^{2}}G_{ab}^{\mu \nu }\sigma _{\mu \nu }\left[
\ln \left( \frac{-x^{2}\Lambda ^{2}}{4}\right) +2\gamma _{E}\right] +\cdots ,
\label{eq:QProp}
\end{eqnarray}%
where $\gamma _{E}\simeq 0.577$ is the Euler constant and $\Lambda $ is the
QCD scale parameter. In Eq. (\ref{eq:QProp}) $G_{ab}^{\alpha \beta
}=G_{A}^{\alpha \beta }t_{ab}^{A}$, where$\ t^{A}=\lambda ^{A}/2$ with $%
\lambda ^{A}$ being the Gell-Mann matrices and $A,B,C=1,\,2\,\ldots 8$. Let
us note that the gluon field strength tensor is fixed at $x=0$, i.e., $%
G_{\alpha \beta }^{A}\equiv G_{\alpha \beta }^{A}(0)$.

To find the required sum rules, we extract the invariant amplitude $\Pi ^{%
\mathrm{OPE}}(p^{2})$ corresponding to the structure $g_{\mu \nu }$, and
equate it to $\Pi ^{\mathrm{Phys}}(p^{2})$. We apply the Borel
transformation to both sides of the obtained equality, which is necessary to
suppress contributions of the higher resonances and continuum states. At the
next stage, using an assumption on quark-hadron duality, we carry out the
continuum subtraction. After these standard manipulations the sum rule
depends on new parameters $M^{2}$ and $s_{0}$: the first of them $M^{2}$ is
the Borel parameter generated by the Borel transformation, whereas $s_{0}$
is the continuum threshold parameter that dissects contributions of the
ground-state and higher resonances from each another. Remaining operations
to find the sum rules for $m_{Y}$ and $f_{Y}$ are similar to ones presented
numerously in the literature, and therefore, we skip further details. It is
worth noting that calculation of $\Pi ^{\mathrm{OPE}}(p^{2})$ in the present
article is performed by taking into account nonperturbative terms up to
dimension 15.

The obtained sum rules contain various vacuum condensates, and depend on the
$s$ quark's mass and on two auxiliary parameters $M^{2}$ and $s_{0}$. Values
of the vacuum condensates and the mass of $s$ quark used in numerical
computations are collected in Table \ref{tab:PM}. Here, we also write down
the parameters of the $\phi $, $f_{0}(980)$, $\eta $, and $\eta ^{\prime }$
mesons which are necessary to calculate partial widths of the decay
processes.
\begin{table}[tbp]
\begin{tabular}{|c|c|}
\hline\hline
Quantity & Value \\ \hline\hline
$\langle \bar{q}q \rangle $ & $-(0.24\pm 0.01)^3~\mathrm{GeV}^3$ \\
$\langle \bar{s}s \rangle $ & $0.8\langle \bar{q}q \rangle$ \\
$m_{0}^2 $ & $(0.8\pm0.1)~\mathrm{GeV}^2$ \\
$\langle \overline{q}g_{s}\sigma Gq\rangle$ & $m_{0}^2\langle \bar{q}q
\rangle $ \\
$\langle \overline{s}g_{s}\sigma Gs\rangle$ & $m_{0}^2\langle \bar{s}s
\rangle $ \\
$\langle\frac{\alpha_sG^2}{\pi}\rangle $ & $(0.012\pm0.004)~\mathrm{GeV}^4$
\\
$m_{s} $ & $93^{+11}_{-5}~\mathrm{MeV} $ \\
$m_{\phi}$ & $(1019.461 \pm 0.019)~\mathrm{MeV}$ \\
$m_{f}$ & $(990 \pm 20)~\mathrm{MeV}$ \\
$m_{\eta}$ & $(547.862\pm 0.018)~\mathrm{MeV}$ \\
$m_{\eta ^{\prime }}$ & $(957.78\pm 0.06)~\mathrm{MeV}$ \\
$f_{\phi }$ & $(215\pm 5)~\mathrm{MeV}$ \\ \hline\hline
\end{tabular}%
\caption{Vacuum condensates and spectroscopic parameters of the mesons used
in numerical computations.}
\label{tab:PM}
\end{table}

The condensates characterize nonperturbative features of the vacuum and do
not depend on a problem under consideration. On the contrary, the Borel and
continuum threshold parameters $M^{2}$ and $s_{0}$ should be chosen for each
sum rule computations individually and must meet restrictions imposed on
them by the QCD sum rule method. The main constraints on $M^{2}$ and $s_{0}$
are connected with convergence of the operator product expansion ($\mathrm{%
OPE}$) which we fix by means of the ratio
\begin{equation}
R(M^{2})=\frac{\Pi ^{\mathrm{DimN}}(M^{2},\ s_{0})}{\Pi (M^{2},\ s_{0})},
\label{eq:Conv}
\end{equation}%
and with the restriction on the pole contribution ($\mathrm{PC}$)
\begin{equation}
\mathrm{PC}=\frac{\Pi (M^{2},\ s_{0})}{\Pi (M^{2},\ \infty )}.  \label{eq:PC}
\end{equation}%
In Eqs.\ (\ref{eq:Conv}) and (\ref{eq:PC}) $\Pi (M^{2},s_{0})$ is the
invariant amplitude $\Pi ^{\mathrm{OPE}}(p^{2})$ obtained after the Borel
transformation and subtraction procedures, and $\Pi ^{\mathrm{DimN}}(M^{2},\
s_{0})$ denotes a last term (or a sum of last few terms) in $\mathrm{OPE}$.
At the minimum of the working window for the Borel parameter, we require $%
R(M^{2})\simeq 0.01$ and use a sum of three terms $\mathrm{DimN}=\mathrm{%
Dim(13+14+15)}$ to estimate $R(M_{\min }^{2})$. At maximum allowed value of $%
M^{2}$, we demand fulfillment of the condition $\mathrm{PC}>0.2.$

In general, $m_{Y}$ and $f_{Y}$ extracted from the sum rules should not
depend on the Borel parameter $M^{2}$. But in actual computations the best thing
one can do is find a plateau where dependence of physical quantities on $%
M^{2}$ is minimal. The continuum threshold parameter $s_{0}$ separates a
ground-state contribution from the ones due to higher resonances and
continuum states. In other words, $s_{0}$ should be below the first excited
state of the particle under discussion $Y$. In the case of conventional
hadrons, masses of excited states are known either from experimental
measurements or from alternative theoretical studies. For exotic particles
the situation is more complicated: there is not  information on their
radial and/or orbital excitations. It is worth noting that for tetraquarks
this problem was addressed only in few publications \cite%
{Maiani:2014,Wang:2014vha,Agaev:2017tzv}. Therefore, one chooses $s_{0}$ by
demanding maximum for $\mathrm{PC}$ and, at the same time, a stability of an
extracting physical quantity. In such analysis very important is control
over self-consistency of the prediction for $m_{Y}$ and $s_{0}$ used for
these purposes: $\sqrt{s_{0}}$ may exceed $m_{Y}$ approximately $[0.3,0.6]~%
\mathrm{MeV}$ to be below a first excited state of $Y$. Uncertainties in the
choice of the $M^{2}$ and $s_{0}$ are main sources of theoretical errors in
the sum rule calculations, which however can be systematically kept under
control.

Numerical analysis allows us to fix the regions
\begin{equation}
M^{2}\in \lbrack 1.2,\ 1.7]~\mathrm{GeV}^{2},\ s_{0}\in \lbrack 6,\ 6.5]~%
\mathrm{GeV}^{2}  \label{eq:Wind1}
\end{equation}%
as ones which obey the constraints imposed on $M^{2}$ and $s_{0}$. Thus, at $%
M^{2}=1.2~\mathrm{GeV}^{2}$ the convergence of the $\mathrm{OPE}$ is
fulfilled, because a contribution of the last three terms to the Borel
transformed and subtracted invariant amplitude $\Pi (M^{2},s_{0})$ does not
exceed $0.3\%$ of its value. At $M^{2}=1.2~\mathrm{GeV}^{2}$ the pole
contribution forms $60\%$ of $\Pi (M^{2},s_{0})$, whereas at $M^{2}=1.7~%
\mathrm{GeV}^{2}$ it amounts to approximately $30\%$ of the whole result.

\begin{figure}[h]
\includegraphics[width=8.8cm]{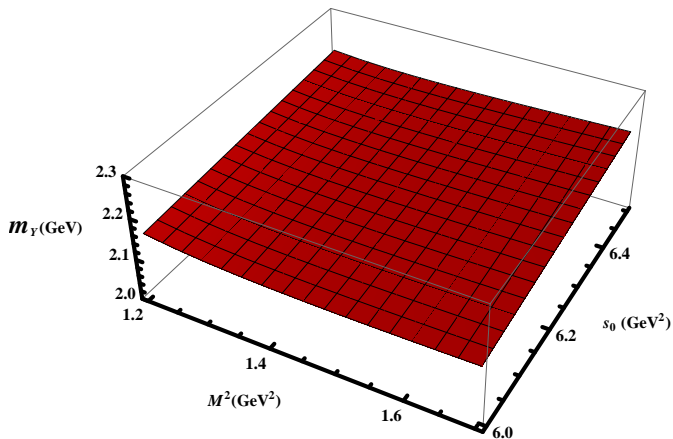}
\caption{The mass $m_{Y}$ of the tetraquark $Y$ as a function of the Borel
and continuum threshold parameters.}
\label{fig:Mass}
\end{figure}
\begin{figure}[h]
\includegraphics[width=8.8cm]{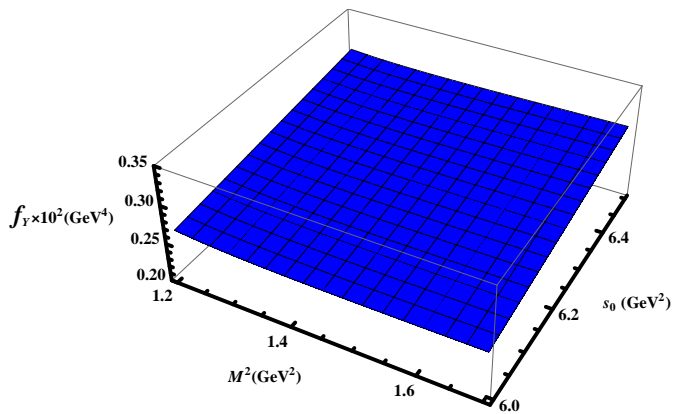}
\caption{Dependence of the coupling $f_{Y}$ on $M^{2}$ and $s_{0}$.}
\label{fig:Coupl}
\end{figure}
The mass $m_{Y}$ and coupling $f_{Y}$ are plotted in Figs.\ \ref{fig:Mass}
and \ref{fig:Coupl} as functions of $M^{2}$ and $s_{0}$: one can inspect
their dependence on the Borel and continuum threshold parameters which is
considerable for $f_{Y}$.

Our results for the spectroscopic parameters of the tetraquark $Y$ are
\begin{eqnarray}
m_{Y} &=&\left( 2173\pm 85\right) ~\mathrm{MeV},\   \notag \\
f_{Y} &=&\left( 2.8\pm 0.5\right) \times 10^{-3}~\mathrm{GeV}^{4}.
\label{eq:Result1}
\end{eqnarray}%
Comparing $m_{Y}$ with $\sqrt{s_{0}}$ we see that $\sqrt{s_{0}}%
-m_{Y}=[0.28,0.38]~\mathrm{MeV}$ is a reasonable mass gap to separate $Y$
from its excitations.

As is seen, the $m_{Y}$ is in excellent agreement with the datum of the
BaBar Collaboration $(2175\pm 10\pm 15)~\mathrm{MeV}$. It is lower than the
new result $(2200\pm 6\pm 5)~\mathrm{MeV}$ reported by BESIII, but within
errors of calculations is compatible with this measurement as well. In this
situation decays of the tetraquark $Y$ become of special interest, because
predictions for partial widths of the different channels and for the full
width of the $Y$ are important to verify our assumption on its structure.

%%%%%%%%%%%%%%%%%%%%%%%%%%%%%%%%%%%%%%%%%%%%%%%%%%%%%%%%%%%%%%%%%%%%%

\section{The decay $Y \to \protect\phi f_{0}(980)$}

\label{sec:Decay1}
%%%%%%%%%%%%%%%%%%%%%%%%%%%%%%%%%%%%%%%%%%%%%%%%%%%%%%%%%%%%%%%%%%%%%%%%%%%%%%
The process $Y\rightarrow \phi f_{0}(980)$ is one of dominant strong decays
of the tetraquark $Y$. To calculate partial width of this channel, we
extract the strong coupling $G_{Y\phi f}$ of the vertex $Y\phi f_{0}(980)$
in the context of the LCSR method and express it in terms of various vacuum
condensates and distribution amplitudes (DAs) of the $\phi $ meson.

To derive the light-cone sum rule for the coupling $G_{Y\phi f}$, we start
from analysis of the correlation function
\begin{equation}
\Pi _{\mu }(p,q)=i\int d^{4}xe^{ipx}\langle \phi (q)|\mathcal{T}%
\{J^{f}(x)J_{\mu }^{Y\dagger }(0)\}|0\rangle ,  \label{eq:CF4}
\end{equation}%
where $J_{\mu }^{Y}(x)$ is the interpolating current of $Y$ \ introduced in
Eq.\ (\ref{eq:Curr}).

As it has been emphasized above, we consider the scalar meson $f_{0}(980)$
[in formulas we use $f=f_{0}(980)$] as a pure $\mathbf{H}$ state.
Interpolating current for such state is given by expression

\begin{eqnarray}
&&J^{f}(x)=\frac{\epsilon \widetilde{\epsilon }}{\sqrt{2}}\left\{ \left[
u_{a}^{T}(x)C\gamma _{5}s_{b}(x)\right] \left[ \overline{u}_{c}(x)\gamma
_{5}C\overline{s}_{e}^{T}(x)\right] \right.  \notag \\
&&\left. +\left[ d_{a}^{T}(x)C\gamma _{5}s_{b}(x)\right] \left[ \overline{d}%
_{c}(x)\gamma _{5}C\overline{s}_{e}^{T}(x)\right] \right\} ,
\label{eq:MixCurr1}
\end{eqnarray}%
where $\epsilon \widetilde{\epsilon }=\epsilon ^{dab}\epsilon ^{dce}$.

Then, the phenomenological side of the sum rule is determined by the formula%
\begin{eqnarray}
&&\Pi _{\mu }^{\mathrm{Phys}}(p,q)=\frac{\langle 0|J^{f}|f\left( p\right)
\rangle }{p^{2}-m_{f}^{2}}\langle f\left( p\right) \phi (q)|Y(p^{\prime
})\rangle  \notag \\
&&\times \frac{\langle Y(p^{\prime })|J_{\mu }^{Y\dagger }|0\rangle }{%
p^{\prime 2}-m_{Y}^{2}}+\cdots ,  \label{eq:Phys1}
\end{eqnarray}%
where $p^{\prime }$, and $p,\ q$ are $4$-momenta of the initial and final
particles, respectively. To simplify $\Pi _{\mu }^{\mathrm{Phys}}(p,q)$ we
express the matrix elements in terms of physical parameters of the particles
involved into the decay process. The matrix element $\langle Y(p^{\prime
})|J_{\mu }^{\dagger }|0\rangle $ is given by Eq.\ (\ref{eq:MElem1}),
whereas for $\langle 0|J^{f}|f\left( p\right) \rangle $ we use
\begin{equation}
\langle 0|J^{f}|f(p)\rangle =F^{f}m_{f}.  \label{eq:Coupl}
\end{equation}%
We parameterize the vertex $\langle f\left( p\right) \phi (q)|Y(p^{\prime
})\rangle $ by means of the expression
\begin{eqnarray}
&&\langle f\left( p\right) \phi (q)|Y(p^{\prime })\rangle =G_{Y\phi f}\left[
(p^{\prime }\cdot q)(\varepsilon ^{\ast }\cdot \varepsilon ^{\prime })\right.
\notag \\
&&\left. -(q\cdot \varepsilon ^{\prime })(p^{\prime }\cdot \varepsilon
^{\ast })\right] ,  \label{eq:MElem3}
\end{eqnarray}%
where $G_{Y\phi f}$ is the strong coupling which should be determined using
the sum rule, and $\varepsilon _{\mu }^{\ast }$ is the polarization vector
of the $\phi $ meson. This information on the matrix elements is enough to
get the phenomenological side of sum rule which reads%
\begin{eqnarray}
&&\Pi _{\mu }^{\mathrm{Phys}}(p,q)=G_{Y\phi f}\frac{m_{Y}f_{Y}m_{f}F_{f}}{%
2\left( p^{\prime 2}-m_{Y}^{2}\right) \left( p^{2}-m_{f}^{2}\right) }  \notag
\\
&&\times \left[ \left( m_{f}^{2}-m_{Y}^{2}-m_{\phi }^{2}\right) \varepsilon
_{\mu }^{\ast }+\frac{m_{Y}^{2}+m_{f}^{2}-m_{\phi }^{2}}{m_{Y}^{2}}p\cdot
\varepsilon ^{\ast }q_{\mu }\right] .  \notag \\
&&  \label{eq:Phys2}
\end{eqnarray}%
It is seen that the function $\Pi _{\mu }^{\mathrm{Phys}}(p,q)$ contains two
Lorentz structures which can be employed to derive the required sum rule. In
the present study we choose the structure proportional to the polarization
vector $\varepsilon _{\mu }^{\ast }$.

The second component of the sum rule $\Pi _{\mu }^{\mathrm{OPE}}(p,q)$ is
obtained by substituting the interpolating currents into the correlation
function (\ref{eq:CF4}), contracting the relevant quark fields, and
expressing a final expression in terms of quarks' light-cone propagators $%
\mathcal{S}_{q}(x)$, and distribution amplitudes of the $\phi $ meson.

After contracting the quark fields the matrix element in Eq.\ (\ref{eq:CF4})
contains numerous terms of the forms
\begin{eqnarray}
&&\left[ A(x)\right] _{\alpha \beta }^{ab}\langle \phi (q)|\overline{s}%
_{\alpha }^{a}(x)s_{\beta }^{b}(0)|0\rangle ,  \notag \\
&&\left[ B(x)\right] _{\alpha \beta }^{ab}\langle \phi (q)|\overline{s}%
_{\alpha }^{a}(0)s_{\beta }^{b}(x)|0\rangle ,  \label{eq:AB}
\end{eqnarray}%
where $\alpha $ and $\beta $ are the spinor indices. Here $A(x)$ and $B(x)$
are some combinations of the propagators $\mathcal{S}_{q}(\pm x)$,$\
\widetilde{\mathcal{S}}_{q}(\pm x)=C\mathcal{S}_{q}^{T}(\pm x)C$, and $%
\gamma _{5(\sigma )}$ matrices. In calculations we use the light-cone
propagator of the $u$, $d$, and $s$ quarks, which is determined by the
formula
\begin{eqnarray}
&&\mathcal{S}_{q}^{ab}(x)=\frac{i\slashed x}{2\pi ^{2}x^{4}}\delta _{ab}-%
\frac{m_{q}}{4\pi ^{2}x^{2}}\delta _{ab}-\frac{\langle \overline{q}q\rangle
}{12}\left( 1-i\frac{m_{q}}{4}\slashed x\right) \delta _{ab}  \notag \\
&&-\frac{x^{2}}{192}m_{0}^{2}\langle \overline{q}q\rangle \left( 1-i\frac{%
m_{q}}{6}\slashed x\right) \delta _{ab}  \notag \\
&&-ig_{s}\int_{0}^{1}du\left\{ \frac{\slashed xG_{ab}^{\mu \nu }(ux)\sigma
_{\mu \nu }}{16\pi ^{2}x^{2}}-\frac{iux_{\mu }}{4\pi ^{2}x^{2}}G_{ab}^{\mu
\nu }(ux)\gamma _{\nu }\right.  \notag \\
&&\left. -\frac{im_{q}}{32\pi ^{2}}G_{ab}^{\mu \nu }(ux)\sigma _{\mu \nu }%
\left[ \ln \left( \frac{-x^{2}\Lambda ^{2}}{4}\right) +2\gamma _{E}\right]
\right\} .  \label{eq:Prop2}
\end{eqnarray}%
The first two terms in (\ref{eq:Prop2}) are the perturbative components of
the propagator, whereas others are nonperturbative contributions. The terms $%
\sim G^{\mu \nu }$ appear due to expansion of $\mathcal{S}_{q}(x)$ on the
light-cone and describe interactions with the gluon field. In our analysis,
we neglect terms proportional to $m_{q}$, but, at the same time, take into
account the ones $\sim m_{s}$.

Apart from propagators the function $\Pi _{\mu }^{\mathrm{OPE}}(p,q)$
depends also on non-local matrix elements of the quark operator $\overline{s}%
s$ sandwiched between the vacuum and $\phi $ state. To express these matrix
elements using the $\phi $ meson's distribution amplitudes, we expand $%
\overline{s}(x)s(0)$ [this analysis is valid for $\overline{s}(0)s(x)$ as
well ] over the full set of Dirac matrices $\Gamma ^{J}$ and project them
onto the color-singlet states
\begin{equation}
\overline{s}_{\alpha }^{a}(x)s_{\beta }^{b}(0)\rightarrow \frac{1}{12}\delta
^{ab}\Gamma _{\beta \alpha }^{J}\left[ \overline{s}(x)\Gamma ^{J}s(0)\right]
,  \label{eq:MatEx}
\end{equation}%
where $\Gamma ^{J}$
\begin{equation}
\Gamma ^{J}=\mathbf{1},\ \gamma _{5},\ \gamma _{\mu },\ i\gamma _{5}\gamma
_{\mu },\ \sigma _{\mu \nu }/\sqrt{2}.
\end{equation}%
The matrix element of the operators $\overline{s}(x)\Gamma ^{J}s(0)$ can be
expanded over $x^{2}$ and written down in terms of the $\phi $ meson's two-
and three-particle DAs of different twist. In the case $\ \Gamma ^{J}=%
\mathbf{1}$ and $\ i\gamma _{\mu }\gamma _{5}$ we use the definitions%
\begin{equation}
\langle 0|\overline{s}(x)s(0)|\phi (q)\rangle =-if_{\phi }^{\perp
}\varepsilon \cdot xm_{\phi }^{2}\int_{0}^{1}due^{i\overline{u}qx}\psi
_{3}^{\parallel }(u),  \label{eq:DA1}
\end{equation}%
and
\begin{eqnarray}
&&\langle 0|\overline{s}(x)\gamma _{\mu }\gamma _{5}s(0)|\phi (q)\rangle =%
\frac{1}{2}f_{\phi }^{\parallel }m_{\phi }\epsilon _{\mu \nu \alpha \beta
}\varepsilon ^{\nu }q^{\alpha }x^{\beta }  \notag \\
&&\times \int_{0}^{1}due^{i\overline{u}qx}\psi _{3}^{\perp }(u).
\label{eq:DA1A}
\end{eqnarray}%
For the structures $\Gamma ^{J}=\gamma _{\mu }\ $and $\sigma _{\mu \nu }$ we
have
\begin{eqnarray}
&&\langle 0|\overline{s}(x)\gamma _{\mu }s(0)|\phi (q)\rangle =f_{\phi
}^{\parallel }m_{\phi }\left\{ \frac{\varepsilon \cdot x}{q\cdot x}q_{\mu
}\right.  \notag \\
&&\times \int_{0}^{1}due^{i\overline{u}qx}\left[ \phi _{2}^{\parallel }(u)+%
\frac{m_{\phi }^{2}x^{2}}{4}\phi _{4}^{\parallel }(u)\right]  \notag \\
&&+\left( \varepsilon _{\mu }-q_{\mu }\frac{\varepsilon \cdot x}{q\cdot x}%
\right) \int_{0}^{1}due^{i\overline{u}qx}\phi _{3}^{\perp }(u)  \notag \\
&&\left. -\frac{1}{2}x_{\mu }\frac{\varepsilon \cdot x}{(q\cdot x)^{2}}%
m_{\phi }^{2}\int_{0}^{1}due^{i\overline{u}qx}C(u)+...\right\} ,
\label{eq:DA2}
\end{eqnarray}%
and
\begin{eqnarray}
&&\langle 0|\overline{s}(x)\sigma _{\mu \nu }s(0)|\phi (q)\rangle =if_{\phi
}^{\perp }\left\{ \left( \varepsilon _{\mu }q_{\nu }-\varepsilon _{\nu
}q_{\mu }\right) \right.  \notag \\
&&\times \int_{0}^{1}due^{i\overline{u}qx}\left[ \phi _{2}^{\perp }(u)+\frac{%
m_{\phi }^{2}x^{2}}{4}\phi _{4}^{\perp }(u)\right]  \notag \\
&&+\frac{1}{2}\left( \varepsilon _{\mu }x_{\nu }-\varepsilon _{\nu }x_{\mu
}\right) \frac{m_{\phi }^{2}}{q\cdot x}\int_{0}^{1}due^{i\overline{u}qx}%
\left[ \psi _{4}^{\perp }(u)-\phi _{2}^{\perp }(u)\right]  \notag \\
&&\left. +\left( q_{\mu }x_{\nu }-q_{\nu }x_{\mu }\right) \frac{\varepsilon
\cdot x}{(q\cdot x)^{2}}m_{\phi }^{2}\int_{0}^{1}due^{i\overline{u}%
qx}D(u)+...\right\} ,  \notag \\
&&{}  \label{eq:DA3}
\end{eqnarray}%
respectively. Here $\overline{u}=1-u$, and $m_{\phi }$ and $\varepsilon $
are the mass and polarization vector of the $\phi $ meson, respectively. In
the equations above the functions $C(u)$ and $D(u)$ denote the combinations
of the two-particle DAs
\begin{eqnarray}
&&C(u)=\psi _{4}^{\parallel }(u)+\phi _{2}^{\parallel }(u)-2\phi _{3}^{\perp
}(u),  \notag \\
&&D(u)=\phi _{3}^{\parallel }(u)-\frac{1}{2}\phi _{2}^{\perp }(u)-\frac{1}{2}%
\psi _{4}^{\perp }(u).  \label{eq:DA4}
\end{eqnarray}%
The twists of the distribution amplitudes are shown as subscripts in the
relevant functions. It is seen that the $C(u)$ and $D(u)$ include the
two-particle leading twist DAs $\phi _{2}^{\parallel (\perp )}(u)$, the
twist-3 distribution amplitudes $\phi _{3}^{\parallel (\perp )}(u)$ and $%
\psi _{3}^{\parallel (\perp )}(u)$, as well as twist-4 distributions $\phi
_{4}^{\parallel (\perp )}(u)$ and $\psi _{4}^{\parallel (\perp )}(u)$.
Expressions of the matrix elements $\langle 0|\overline{s}(x)\Gamma
^{J}G_{\mu \nu }(vx)s(0)|\phi (q)\rangle $ in terms of the higher twist DAs
of the $\phi $ meson, as well as detailed information on their properties
were reported in Refs.\ \cite%
{Ball:1996tb,Ball:1998sk,Ball:1998ff,Ball:2007rt,Ball:2007zt}.

The main contribution to $\Pi _{\mu }^{\mathrm{OPE}}(p,q)$ comes from the
terms (\ref{eq:AB}), where all of the propagators are replaced by their
perturbative components (see, Fig.\ \ref{fig:Diag1}). Contribution of this
diagram can be computed using the $\phi $ meson two-particle distribution
amplitudes. The one gluon-exchange diagrams shown in Fig.\ \ref{fig:Diag2}
are corrections, which can be expressed and calculated by utilizing
three-particle DAs of the $\phi $ meson. An analytic expression of the $\Pi
_{\mu }^{\mathrm{OPE}}(p,q)$ in terms of the $\phi $ meson's DAs is rather
cumbersome, therefore we do not provide it here.

\begin{widetext}

\begin{figure}[h]
\includegraphics[width=7.8cm]{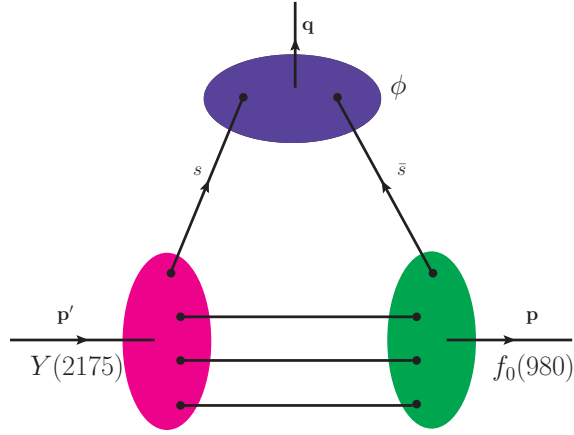}
\caption{The leading order diagram contributing to $\Pi _{\protect\mu }^{%
\mathrm{OPE}}(p,q)$.}
\label{fig:Diag1}
\end{figure}
\begin{figure}[h]
\includegraphics[width=14.8cm]{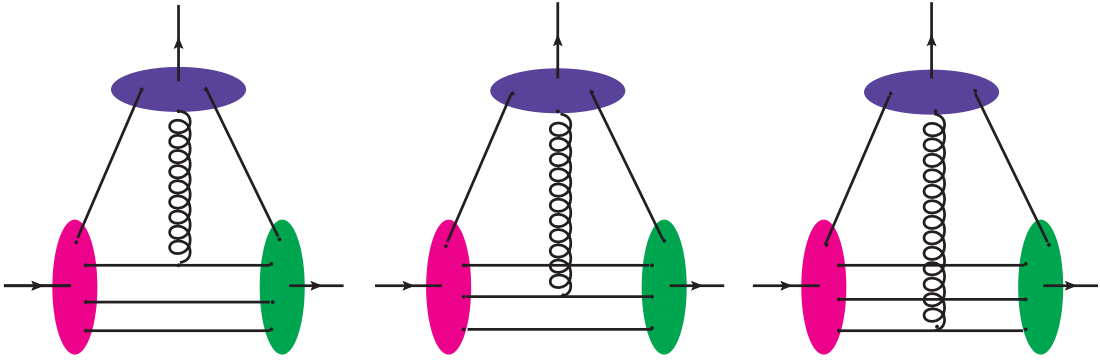}
\caption{The one-gluon exchange diagrams, which can be expressed in terms of
the $\protect\phi$ meson's three-particle DAs.}
\label{fig:Diag2}
\end{figure}

\end{widetext}

In our analysis we employ the invariant amplitude $\Pi ^{\mathrm{OPE}%
}(p^{\prime 2},p^{2})$ proportional to $\varepsilon _{\mu }^{\ast }$ and
match it to corresponding function from $\Pi _{\mu }^{\mathrm{Phys}}(p,q)$.
These amplitudes depend on $p^{\prime 2}$ and $p^{2}$, therefore one should
perform the double Borel transformation over the variables $p^{\prime 2}$
and $p^{2}$
\begin{equation}
\Pi ^{\mathrm{OPE}}(M_{1}^{2},M_{2}^{2})=\mathcal{B}_{p^{\prime
2}}^{M_{1}^{2}}\mathcal{B}_{p^{2}}^{M_{2}^{2}}\Pi ^{\mathrm{OPE}}(p^{\prime
2},p^{2}).
\end{equation}%
The Borel transformed amplitude $\Pi ^{\mathrm{OPE}}\left( M_{1}^{2},\
M_{2}^{2}\right) $ can be calculated in accordance with a scheme explained
in Ref.\ \cite{Agaev:2016srl}, and expressed as a double dispersion
integral. But to simplify manipulations following after the Borel
transformation, we can relate the parameters $M_{1}^{2}$ and $M_{2}^{2}$ to
each other using $\frac{M_{1}^{2}}{M_{2}^{2}}=\frac{m_{Y}^{2}}{m_{f}^{2}}$
and introduce a common parameter $M^{2}$ through the relation
\begin{equation}
\frac{1}{M^{2}}=\frac{1}{M_{1}^{2}}+\frac{1}{M_{2}^{2}}.
\end{equation}%
This implies replacements
\begin{eqnarray}
M_{1}^{2} &=&\frac{m_{f}^{2}+m_{Y}^{2}}{m_{f}^{2}}M^{2},  \notag \\
M_{2}^{2} &=&\frac{m_{f}^{2}+m_{Y}^{2}}{m_{Y}^{2}}M^{2},
\end{eqnarray}%
in the sum rules, and allows us to perform integration over one of variables
in the double dispersion integral. The obtained expressions in this step
depend also on the parameter $u_{0}$ with
\begin{equation}
u_{0}=\frac{M_{1}^{2}}{M_{1}^{2}+M_{2}^{2}}=\frac{m_{Y}^{2}}{%
m_{f}^{2}+m_{Y}^{2}}.
\end{equation}%
As a result of the above procedure we get a single integral representation
for $\Pi ^{\mathrm{OPE}}\left( M^{2}\right) $ which considerably simplifies
the continuum subtraction: formulas necessary to carry out this procedure
can be found in Appendix B of Ref.\ \cite{Agaev:2016srl}.

The DAs of the $\phi $ meson depend on numerous parameters. For example, the
leading twist DAs of the longitudinally and transversely polarized $\phi $
meson are given by the expression
\begin{equation}
\phi _{2}^{\parallel (\perp )}(u)=6u\overline{u}\left[ 1+\sum_{n=2}^{\infty
}a_{n}^{\parallel (\perp )}C_{n}^{3/2}(2u-1)\right] ,  \label{eq:LTDA}
\end{equation}%
where $C_{n}^{m}(2u-1)$ are the Gegenbauer polynomials. Equation (\ref%
{eq:LTDA}) is the general expression for $\phi _{2}^{\parallel (\perp )}(u)$%
. In our calculations we employ twist-2 DAs with a non-asymptotic term $%
a_{2}^{\parallel (\perp )}\neq 0$. The models for the higher twist DAs, and
values of the relevant parameters at the normalization scale $\mu _{0}=1\,%
\mathrm{GeV}$ are taken from Refs.\ \cite{Ball:2007rt,Ball:2007zt} (see
Tables 1 and 2 in Ref.\ \cite{Ball:2007zt}).

The sum rule for the coupling $G_{Y\phi f}$ contains the quark, gluon and
mixed condensates and the $s$-quark mass which are moved to Table \ref%
{tab:PM}. The spectroscopic parameters of the particles involved into the
decay $Y\rightarrow \phi f_{0}(980)$ are also input information of
computations. The mass and coupling of the tetraquark $Y$ have been
evaluated in the present work. For the mass of the $\phi $ and $f_{0}(980)$
mesons we use their experimental values (see Table \ref{tab:PM}). The
coupling $F_{f}$ of the meson $f_{0}(980)$ is borrowed from Ref.\ \cite%
{Agaev:2017cfz}, where it was treated as the four-quark system
\begin{equation}
F_{f}\equiv F_{\mathbf{H}}=\left( 1.35\pm 0.34\right) \times 10^{-3}~\mathrm{%
GeV}^{4}.
\end{equation}%
Finally, the sum rule depends on the Borel and continuum threshold
parameters: $M^{2}$ and $s_{0}$ are auxiliary parameters of computations,
and the result should be insensitive to their choices. But in real analysis
we can only minimize these effects and fix convenient working windows for
the $M^{2} $ and $s_{0}$
\begin{equation}
M^{2}\in \lbrack 2.4,\ 3.4]~\mathrm{GeV}^{2},\ s_{0}\in \lbrack 6,\ 6.5]~%
\mathrm{GeV}^{2}.  \label{eq:Wind2}
\end{equation}

In accordance with our studies the strong coupling $G_{Y\phi f}$ is equal to
\begin{equation}
G_{Y\phi f}=(1.62\pm 0.41)\ \mathrm{GeV}^{-1}.
\end{equation}%
The width of the decay $Y\rightarrow \phi f_{0}(980)$ is determined by the
expression%
\begin{equation}
\Gamma (Y\rightarrow \phi f)=\frac{G_{Y\phi f}^{2}m_{\phi }^{2}}{24\pi }%
\lambda \left( 3+\frac{2\lambda ^{2}}{m_{\phi }^{2}}\right) ,
\label{eq:DW0A}
\end{equation}%
where
\begin{eqnarray}
\lambda &\equiv &\lambda (m_{Y},m_{\phi },m_{f})=\frac{1}{2m_{Y}}\left[
m_{Y}^{4}+m_{\phi }^{4}+m_{f}^{4}\right.  \notag \\
&&\left. -2\left( m_{Y}^{2}m_{\phi }^{2}+m_{Y}^{2}m_{f}^{2}+m_{\phi
}^{2}m_{f}^{2}\right) \right] ^{1/2}.  \label{eq:Lambda}
\end{eqnarray}%
Then computations yield
\begin{equation}
\Gamma (Y\rightarrow \phi f)=(49.2\pm 17.6)~\mathrm{MeV}.  \label{eq:DW0}
\end{equation}%
The prediction for $\Gamma (Y\rightarrow \phi f)$ is the main result of this
section which will be used to estimate the full width of the tetraquark $Y$.

%%%%%%%%%%%%%%%%%%%%%%%%%%%%%%%%%%%%%%%%%%%%%%%%%%%%%%%%%%%%%%%%%%%%%%%%%%%%%%

\section{The decays $Y\rightarrow $ $\protect\phi \protect\eta $ and $%
Y\rightarrow $ $\protect\phi \protect\eta ^{\prime }$}

\label{sec:Decay2}
%%%%%%%%%%%%%%%%%%%%%%%%%%%%%%%%%%%%%%%%%%%%%%%%%%%%%%%%%%%%%%%%%%%%%%%%%%%%%%

The next two strong decays of the tetraquark $Y$ are the channels $%
Y\rightarrow $ $\phi \eta $ and $Y\rightarrow $ $\phi \eta ^{\prime }$.
Here, we consider in a detailed form the dominant process $Y\rightarrow $ $%
\phi \eta $, and write down final results for the second mode $Y\rightarrow $
$\phi \eta ^{\prime }$.

In the framework of the LCSR method the correlation function necessary to
study the vertex $Y\phi \eta $ is given by the expression
\begin{equation}
\Pi _{\mu \nu }(p,q)=i\int d^{4}xe^{ipx}\langle \eta (q)|\mathcal{T}\{J_{\mu
}^{\phi }(x)J_{\nu }^{Y\dagger }(0)\}|0\rangle ,  \label{eq:CF5}
\end{equation}%
where $J_{\mu }^{\phi }(x)$ is the interpolating current for the vector $%
\phi $ meson%
\begin{equation}
J_{\mu }^{\phi }(x)=\overline{s}(x)\gamma _{\mu }s(x).  \label{eq:Curr1}
\end{equation}%
The phenomenological side of the sum rule can be written down in the form
\begin{eqnarray}
&&\Pi _{\mu \nu }^{\mathrm{Phys}}(p,q)=\frac{\langle 0|J_{\mu }^{\phi }|\phi
\left( p\right) \rangle }{p^{2}-m_{\phi }^{2}}\langle \phi \left( p\right)
\eta (q)|Y(p^{\prime })\rangle  \notag \\
&&\times \frac{\langle Y(p^{\prime })|J_{\nu }^{Y\dagger }|0\rangle }{%
p^{\prime 2}-m_{Y}^{2}}+\cdots ,  \label{eq:Phys3}
\end{eqnarray}%
and simplified further using the matrix elements:
\begin{equation}
\langle 0|J_{\mu }^{\phi }|\phi \left( p\right) \rangle =f_{\phi }m_{\phi
}\epsilon _{\mu },  \label{eq:ME2}
\end{equation}%
and
\begin{equation}
\langle \phi \left( p\right) \eta (q)|Y(p^{\prime })\rangle =g_{Y\phi \eta
}\varepsilon _{\mu \nu \alpha \beta }p^{\mu }q^{\nu }\epsilon ^{\ast \alpha
}\epsilon ^{\prime \beta },  \label{eq:ME3}
\end{equation}%
where $\epsilon ^{\prime \beta }$ is the polarization vector of the
tetraquark $Y$, and $g_{Y\phi \eta }$ is the strong coupling corresponding
to the vertex $Y\phi \eta $.

Simple manipulations allow us to recast $\Pi _{\mu \nu }^{\mathrm{Phys}%
}(p,q) $ into the form%
\begin{eqnarray}
&&\Pi _{\mu \nu }^{\mathrm{Phys}}(p,q)=g_{Y\phi \eta }\frac{f_{\phi }m_{\phi
}f_{Y}m_{Y}}{\left( p^{2}-m_{\phi }^{2}\right) \left( p^{\prime
2}-m_{Y}^{2}\right) }  \notag \\
&&\times \varepsilon _{\mu \nu \alpha \beta }p^{\alpha }q^{\beta }+\cdots ,
\label{eq:Phys4}
\end{eqnarray}%
where the only term is the contribution arising from the ground-state
particles: effects of the higher resonances and continuum states are denoted
by dots. The correlation function $\Pi _{\mu \nu }^{\mathrm{Phys}}(p,q)$ has
a simple Lorentz structure. The invariant amplitude $\Pi ^{\mathrm{Phys}%
}(p^{\prime 2},p^{2})$, which will be used to derive the sum rule for the
coupling $g_{Y\phi \eta }$, can be obtained from Eq.\ (\ref{eq:Phys4}) by
factoring out the structure $\varepsilon _{\mu \nu \alpha \beta }p^{\alpha
}q^{\beta }$.

We extract the second component of the sum rule, i. e., the invariant
amplitude $\Pi ^{\mathrm{OPE}}(p^{\prime 2},p^{2})$ from the correlation
function $\Pi _{\mu \nu }^{\mathrm{OPE}}(p,q)$. In the case under analysis
it is given by the following expression
\begin{eqnarray}
&&\Pi _{\mu \nu }^{\mathrm{OPE}}(p,q)=-i\int d^{4}xe^{ipx}\left[ \gamma _{5}%
\widetilde{S}_{s}^{ib}(x){}\gamma _{\mu }\widetilde{S}_{s}^{bi}(-x)\gamma
_{5}\gamma _{\nu }\right.  \notag \\
&&\left. +\gamma _{\nu }\gamma _{5}\widetilde{S}_{s}^{ib}(x){}\gamma _{\mu }%
\widetilde{S}_{s}^{bi}(-x)\gamma _{5}\right] _{\alpha \beta }\langle \eta
(q)|\overline{u}_{\alpha }^{a}(0)u_{\beta }^{a}(0)|0\rangle .  \notag \\
&&  \label{eq:CF6}
\end{eqnarray}

As is seen, the correlation function is written down in terms of the $s$
quark propagators and local matrix elements of the $\eta $ meson. Dependence
of $\Pi _{\mu \nu }^{\mathrm{OPE}}(p,q)$ on the local matrix elements of a
final meson is the distinctive feature of the LCSR method when applied to
tetraquark-meson-meson vertices. Treatment of such vertices requires some
additional manipulations, which we are going to explain below. But before
that we have to find $\Pi _{\mu \nu }^{\mathrm{OPE}}(p,q)$ by rewriting the
matrix elements $\langle \eta (q)|\overline{u}_{\alpha }^{a}(0)u_{\beta
}^{a}(0)|0\rangle $ in terms of the $\eta $ meson's parameters. To this end,
we expand $\overline{u}_{\alpha }^{a}(0)u_{\beta }^{a}(0)$ and determine the
standard matrix elements of the $\eta $ meson that contribute to the
correlation function. These operations have been discussed in the previous
section, therefore here we omit further details.

Performed analysis shows that the matrix element $\langle \eta (q)|\overline{%
u}\gamma _{\mu }\gamma _{5}u|0\rangle $ contributes to the correlation
function $\Pi _{\mu \nu }^{\mathrm{OPE}}(p,q)$. It is defined by the formula
\begin{equation}
\langle \eta (q)|\overline{u}\gamma _{\mu }\gamma _{5}u|0\rangle =-i\frac{%
f_{\eta }^{q}}{\sqrt{2}}q_{\mu },  \label{eq:ME}
\end{equation}%
where $f_{\eta }^{q}$ is the decay constant of the $\eta $ meson's $"q"$
component. The matrix element (\ref{eq:ME}) differs from similar expressions
of other pseudoscalar mesons. This is connected with the mixing in the $\eta
-\eta ^{\prime }$ system which can be described using either the
octet-singlet or quark-flavor basis of the flavor $SU_{f}(3)$ group. The
latter is more convenient and simple for applications, and was used in
Refs.\ \cite{Agaev:2014wna,Agaev:2015faa,Agaev:2016dsg} to explore different
exclusive processes. This scheme is utilized in the present work as well.

In the quark-flavor basis the decay constants of the mesons $\eta $ and $%
\eta ^{\prime }$ can be extracted from the equality%
\begin{equation}
\left(
\begin{array}{cc}
f_{\eta }^{q} & f_{\eta }^{s} \\
f_{\eta ^{\prime }}^{q} & f_{\eta ^{\prime }}^{s}%
\end{array}%
\right) =U(\varphi )\left(
\begin{array}{cc}
f_{q} & 0 \\
0 & f_{s}%
\end{array}%
\right) ,  \label{eq:Couplings}
\end{equation}%
where $U(\varphi )$ is the mixing matrix
\begin{equation}
U(\varphi )=\left(
\begin{array}{cc}
\cos \varphi & -\sin \varphi \\
\sin \varphi & \cos \varphi%
\end{array}%
\right) ,  \label{eq:MixingM}
\end{equation}%
with  $\varphi
=39^{\circ }.3\pm 1^{\circ }.0$ being the mixing angle in the quark-flavor
basis. The constants $f_{q}$
and $f_{s}$ in Eq.\ (\ref{eq:Couplings})   are given by the formulas
\begin{equation}
f_{q}=(1.07\pm 0.02)f_{\pi },\ \ f_{s}=(1.34\pm 0.06)f_{\pi },
\end{equation}%
where $f_{\pi }=131~\mathrm{MeV}$ is the pion decay constant.

Using Eqs.\ (\ref{eq:CF6}) and (\ref{eq:ME}), we can obtain the invariant
amplitude $\Pi ^{\mathrm{OPE}}(p^{\prime 2},p^{2})$ which should be equated
to $\Pi ^{\mathrm{Phys}}(p^{\prime 2},p^{2})$ in order to derive the sum
rule for the strong coupling $g_{Y\phi \eta }$. But, as we have been
emphasized above, a treatment of tetraquark-meson-meson vertices in the
context of the LCSR method differs from standard analysis of the previous
section \cite{Agaev:2016dev}. In fact, the LCSR for vertices of conventional
mesons depends on distribution amplitudes of one of final mesons, which
contain all information about nonperturbative dynamical features of the
meson. The same arguments are valid for the tetraquark-tetraquark-meson
vertices as well \cite{Agaev:2016srl}. But in the case of the
tetraquark-meson-meson vertices, after contracting relevant quark fields,
due to a four-quark structure of the tetraquark, the correlation function
contains only local matrix elements of one of final mesons. Then
the momentum of this meson should be set $q=0$ which is necessary to satisfy
the four-momentum conservation at the vertex. This leads to essential
modifications in the calculational scheme, because now we have to complete
the LCSR method with technical methods of the soft-meson approximation \cite%
{Belyaev:1994zk,Agaev:2016dev}.

In the soft limit $q\rightarrow 0$, we get $p^{\prime }=p$, as a result we
have to perform one-variable Borel transformation of the invariant
amplitudes \cite{Agaev:2016dev}. For the physical (phenomenological) side
this leads to the formula
\begin{equation}
\mathcal{B}\Pi ^{\mathrm{Phys}}(p^{2})=g_{Y\phi \eta }f_{\phi }m_{\phi
}f_{Y}m_{Y}\frac{e^{-m^{2}/M^{2}}}{M^{2}}+\cdots ,  \label{eq:CF5B}
\end{equation}%
where $m^{2}=(m_{\phi }^{2}+m_{Y}^{2})/2.$

In the soft-meson approximation the phenomenological side of the sum rule
has a more complicated organization than in the case of full LCSR method.
The reason is that in the soft limit contributions connected with higher
resonances and continuum states demonstrate complicated behavior. Indeed,
some of these terms even after the Borel transformation remain unsuppressed
and appear as contaminations in the phenomenological side \cite%
{Belyaev:1994zk}. Therefore, before carrying out the continuum subtraction
they should be excluded from $\mathcal{B}\Pi ^{\mathrm{Phys}}(p^{2})$ by
means of some manipulations. This problem is solved by acting on the
phenomenological side of the sum rule by the operator \cite%
{Belyaev:1994zk,Ioffe:1983ju}
\begin{equation}
\mathcal{P}(M^{2},m^{2})=\left( 1-M^{2}\frac{d}{dM^{2}}\right)
M^{2}e^{m^{2}/M^{2}},
\end{equation}%
that eliminates contaminating terms. Then contributions of higher resonances
with regular behavior can be subtracted from the QCD side by benefiting from
the quark-hadron duality assumption.

The operator $\mathcal{P}(M^{2},m^{2})$ should also be applied to the QCD
side of the sum rule. Then the strong coupling $g_{Y\phi \eta }$ can be
determined from the sum rule
\begin{equation}
g_{Y\phi \eta }=\frac{1}{f_{\phi }m_{\phi }f_{Y}m_{Y}}\mathcal{P}%
(M^{2},m^{2})\Pi ^{\mathrm{OPE}}\left( M^{2},s_{0}\right) ,
\label{eq:Coupl1}
\end{equation}%
where $\Pi ^{\mathrm{OPE}}(M^{2},s_{0})$ is the invariant amplitude $\Pi ^{%
\mathrm{OPE}}(p^{2})$ after the Borel transformation and continuum
subtraction procedures. Our calculations carried out by taking into account
nonperturbative terms up to dimension 5 yield
\begin{eqnarray}
&&\Pi ^{\mathrm{OPE}}\left( M^{2},s_{0}\right) =\frac{f_{\eta }^{q}m_{s}}{8%
\sqrt{2}\pi ^{2}}\int_{4m_{s}^{2}}^{s_{0}}dse^{-s/M^{2}}  \notag \\
&&+\frac{f_{\eta }^{q}m_{s}^{2}}{6\sqrt{2}M^{2}}\langle \overline{s}s\rangle
+\frac{f_{\eta }^{q}}{12\sqrt{2}M^{2}}\langle \overline{s}g_{s}\sigma
Gs\rangle .  \label{eq:CF7}
\end{eqnarray}

The width of the decay $Y\rightarrow $ $\phi \eta $ is given by the
following expression
\begin{equation}
\Gamma (Y\rightarrow \phi \eta )=\frac{g_{Y\phi \eta }^{2}\lambda
^{3}(m_{Y},m_{\phi },m_{\eta })}{12\pi },  \label{eq:DWGen}
\end{equation}%
Numerical analysis leads to the results%
\begin{eqnarray}
&&g_{Y\phi \eta }=(1.85\pm 0.38)~\mathrm{GeV}^{-1},  \notag \\
&&\Gamma (Y\rightarrow \phi \eta )=(35.8\pm 10.4)~\mathrm{MeV}.
\label{eq:DW1}
\end{eqnarray}%
It is worth noting that in computations of $g_{Y\phi \eta },$ we have used
the following working regions for $M^{2}$ and $s_{0}$
\begin{equation}
M^{2}\in \lbrack 1.3,\ 1.8]~\mathrm{GeV}^{2},\ s_{0}\in \lbrack 6,\ 6.5]~%
\mathrm{GeV}^{2}.  \label{eq:Wind3}
\end{equation}

The partial width of the second process $Y\rightarrow \phi \eta ^{\prime }$
can be computed by utilizing the expressions obtained for the first decay.
The corrections are connected with mass of the $\eta ^{\prime }$ meson and
coupling $f_{\eta ^{\prime }}^{q}$, and required replacements
\begin{equation}
f_{\eta ^{\prime }}^{q}=f_{q}\sin \varphi ,\ \lambda \rightarrow \lambda
(m_{Y},m_{\phi },m_{\eta ^{\prime }}),
\end{equation}%
can be easily implemented into analysis. For the parameters of the second
process we obtain
\begin{eqnarray}
g_{Y\phi \eta ^{\prime }} &=&(1.59\pm 0.31)~\mathrm{GeV}^{-1},  \notag \\
\Gamma (Y &\rightarrow &\phi \eta ^{\prime })=(6.1\pm 1.7)~\mathrm{MeV}.
\label{eq:DW2}
\end{eqnarray}

Saturating the full width of the $Y$ resonance by three decay channels
considered in the present work, we get%
\begin{equation}
\Gamma _{\mathrm{full}}=(91.1\pm 20.5)~\mathrm{MeV}.  \label{eq:FullW}
\end{equation}%
This estimate coincides neither with the BaBar data nor with measurements of
the BESIII Collaboration, but is close to the latter.

%%%%%%%%%%%%%%%%%%%%%%%%%%%%%%%%%%%%%%%%%%%%%%%%%%%%%%%%%%%%%%%%%%%%%

\section{Analysis and concluding notes}

\label{sec:Conclusions}
%%%%%%%%%%%%%%%%%%%%%%%%%%%%%%%%%%%%%%%%%%%%%%%%%%%%%%%%%%%%%%%%%%%%%%%%%%%%%%
We have explored the resonance $Y$ by modeling it as a light vector
tetraquark with the content $[su][\overline{s}\overline{u}]$. In the
tetraquark model it was considered until now as a vector $[ss][\overline{s}%
\overline{s}]$ or $(s\overline{s})(s\overline{s})$ particles. Our treatment
is motivated by the dominant decay channel $Y\rightarrow $ $\phi f_{0}(980)$
of the $Y$, where it was observed as a resonant structure in the $\phi
f_{0}(980)$ invariant mass distribution. A suggestion on the quark content
of the $Y$ depends on the structures of the final-state particles: one can
consider the $f_{0}(980)$ either as a scalar meson $\overline{s}s$ or as a
particle composed of the four valence quarks. In the second picture the
vector compound $Y=[su][\overline{s}\overline{u}]$ emerges as a quite
natural assignment for this resonance. Calculations carried out in the
present work lead to the following predictions for $m_{Y}$ and $\Gamma _{%
\mathrm{full}}$ of such a state
\begin{eqnarray}
m_{Y} &=&(2173\pm 85)~\mathrm{MeV},\ \Gamma _{\mathrm{full}}=(91.1\pm 20.5)~%
\mathrm{MeV}.  \notag \\
&&  \label{eq:Discussion1}
\end{eqnarray}

The result for the mass $m_{Y}$ is in accord with the BaBar data, but is
compatible with BESIII measurements as well. The full width $\Gamma _{%
\mathrm{full}}$ has the small overlapping region with $\Gamma =(58\pm 16\pm
20)~\mathrm{MeV}$ extracted in Ref.\ \cite{Aubert:2006bu}, but agreement
with data of the BESIII Collaboration is considerably better. In
calculations of the $\Gamma _{\mathrm{full}}$, we have taken into account
only three strong decays of the resonance $Y$. But decay modes $Y\rightarrow
\phi \pi \pi ,\ K^{+}K^{-}\pi ^{+}\pi ^{-},\ K^{\ast }(892)^{0}\overline{K}%
^{\ast }(892)^{0}$ of $Y$ (seen experimentally and/or theoretically
possible) and other channels have not been included into analysis. Partial
width of these decays may significantly improve the present prediction for $%
\Gamma _{\mathrm{full}}$.

Encouraging is our estimate for the ratio
\begin{equation}
\frac{\Gamma (Y\rightarrow \phi \eta )}{\Gamma (Y\rightarrow \phi f)}\approx
0.73,
\end{equation}%
which almost coincides with its experimental value $\approx 0.74$. The
latter has been extracted from  available information on the ratios \cite%
{Tanabashi:2018oca} [$Y$ is denoted there $\phi (2170)$]%
\begin{equation}
\frac{\Gamma (Y\rightarrow \phi \eta )\times \Gamma (Y\rightarrow e^{+}e^{-})%
}{\Gamma _{\mathrm{total}}}=1.7\pm 0.7\pm 1.3,
\end{equation}%
and%
\begin{equation}
\frac{\Gamma (Y\rightarrow \phi f)\times \Gamma (Y\rightarrow e^{+}e^{-})}{%
\Gamma _{\mathrm{total}}}=2.3\pm 0.3\pm 0.3.
\end{equation}%
Unfortunately, precision of the experimental data and uncertainties of the
theoretical results do not allow us to make more strong statements about
decay modes of the tetraquark $Y$.

As is seen, our suggestion on a nature of the resonance $Y(2175)$ as the
vector tetraquark with the content $[su][\overline{s}\overline{u}]$ has led
to reasonable agreements with existing experimental data. Theoretical
analyses of decay channels left beyond the scope of the present work, as
well as their detailed experimental studies will be of great help to answer
open questions about the structure of the resonance $Y(2175)$.


\begin{thebibliography}{99}
%\cite{Aubert:2006bu}

\bibitem{Aubert:2006bu} B.~Aubert \textit{et al.} [BaBar Collaboration],
%``A Structure at 2175-MeV in $e^{+} e^{-} \to \phi$ f0(980) Observed via Initial-State Radiation,''
Phys.\ Rev.\ D \textbf{74}, 091103 (2006).
%doi:10.1103/PhysRevD.74.091103 [hep-ex/0610018].
%%CITATION = doi:10.1103/PhysRevD.74.091103;%%
%160 citations counted in INSPIRE as of 27 May 2019

%\cite{Ablikim:2014pfc}

\bibitem{Ablikim:2014pfc} M.~Ablikim \textit{et al.} [BESIII Collaboration],
%``Study of $J/\psi \to \eta \phi \pi^+ \pi^-$ at BESIII,''
Phys.\ Rev.\ D \textbf{91}, 052017 (2015).
%doi:10.1103/PhysRevD.91.052017 [arXiv:1412.5258 [hep-ex]].
%%CITATION = doi:10.1103/PhysRevD.91.052017;%%
%30 citations counted in INSPIRE as of 27 May 2019

%\cite{Ablikim:2019tpp}

\bibitem{Ablikim:2019tpp} M.~Ablikim \textit{et al.} [BESIII Collaboration],
%``Cross section measurements of $e^{+}e^{-} \to K^{+}K^{-}K^{+}K^{-} $ and $ \phi K^{+}K^{-}$ at center-of-mass energies from 2.10 to 3.08 GeV,''
Phys.\ Rev.\ D \textbf{100}, 032009 (2019).
%doi:10.1103/PhysRevD.100.032009 [arXiv:1907.06015 [hep-ex]].
%%CITATION = doi:10.1103/PhysRevD.100.032009;%%
%1 citations counted in INSPIRE as of 11 Jan 2020

%\cite{Ablikim:2007ab}

\bibitem{Ablikim:2007ab} M.~Ablikim \textit{et al.} [BES Collaboration],
%``Observation of Y(2175) in J / psi ---> eta phi f(0)(980),''
Phys.\ Rev.\ Lett.\ \textbf{100}, 102003 (2008).
%doi:10.1103/PhysRevLett.100.102003 [arXiv:0712.1143 [hep-ex]].
%%CITATION = doi:10.1103/PhysRevLett.100.102003;%%
%111 citations counted in INSPIRE as of 27 May 2019

%\cite{Shen:2009zze}

\bibitem{Shen:2009zze} C.~P.~Shen \textit{et al.} [Belle Collaboration],
%``Observation of the phi(1680) and the Y(2175) in e+e- ---> phi pi+ pi-,''
Phys.\ Rev.\ D \textbf{80}, 031101 (2009).
%doi:10.1103/PhysRevD.80.031101 [arXiv:0808.0006 [hep-ex]].
%%CITATION = doi:10.1103/PhysRevD.80.031101;%%
%86 citations counted in INSPIRE as of 27 May 2019

%\cite{Ablikim:2018iyx}

\bibitem{Ablikim:2018iyx} M.~Ablikim \textit{et al.} [BESIII Collaboration],
%``Measurement of $e^{+} e^{-} \rightarrow K^{+} K^{-}$ cross section at $\sqrt{s} = 2.00 - 3.08$ GeV,''
Phys.\ Rev.\ D \textbf{99}, 032001 (2019).
%doi:10.1103/PhysRevD.99.032001 [arXiv:1811.08742 [hep-ex]].
%%CITATION = doi:10.1103/PhysRevD.99.032001;%%
%2 citations counted in INSPIRE as of 28 May 2019

%\cite{Ablikim:2018xuz}

\bibitem{Ablikim:2018xuz} M.~Ablikim \textit{et al.} [BESIII Collaboration],
%``Observation and study of $J/\psi\rightarrow\phi\eta\eta'$ at BESIII,''
arXiv:1901.00085 [hep-ex]. %%CITATION = ARXIV:1901.00085;%%
%7 citations counted in INSPIRE as of 24 May 2019

%\cite{Ding:2007pc}

\bibitem{Ding:2007pc} G.~J.~Ding and M.~L.~Yan,
%``Y(2175): Distinguish Hybrid State from Higher Quarkonium,''
Phys.\ Lett.\ B \textbf{657}, 49 (2007).
%doi:10.1016/j.physletb.2007.10.020 [hep-ph/0701047].
%%CITATION = doi:10.1016/j.physletb.2007.10.020;%%
%71 citations counted in INSPIRE as of 28 May 2019

%\cite{Wang:2012wa}

\bibitem{Wang:2012wa} X.~Wang, Z.~F.~Sun, D.~Y.~Chen, X.~Liu and T.~Matsuki,
%``Non-strange partner of strangeonium-like state Y(2175),''
Phys.\ Rev.\ D \textbf{85}, 074024 (2012).
%doi:10.1103/PhysRevD.85.074024 [arXiv:1202.4139 [hep-ph]].
%%CITATION = doi:10.1103/PhysRevD.85.074024;%%23 citations counted in INSPIRE as of 28 May 2019

%\cite{MartinezTorres:2008gy}

\bibitem{MartinezTorres:2008gy} A.~Martinez Torres, K.~P.~Khemchandani,
L.~S.~Geng, M.~Napsuciale and E.~Oset,
%``The X(2175) as a resonant state of the phi K anti-K system,''
Phys.\ Rev.\ D \textbf{78}, 074031 (2008).
%doi:10.1103/PhysRevD.78.074031 [arXiv:0801.3635 [nucl-th]].
%%CITATION = doi:10.1103/PhysRevD.78.074031;%% %%105 citations counted in INSPIRE as of 28 May 2019

%\cite{AlvarezRuso:2009xn}

\bibitem{AlvarezRuso:2009xn} L.~Alvarez-Ruso, J.~A.~Oller and J.~M.~Alarcon,
%``On the phi(1020) f0(980) S-wave scattering and the Y(2175) resonance,''
Phys.\ Rev.\ D \textbf{80}, 054011 (2009).
%%doi:10.1103/PhysRevD.80.054011 [arXiv:0906.0222 [hep-ph]].
%%CITATION = doi:10.1103/PhysRevD.80.054011;%%
%37 citations counted in INSPIRE as of 28 May

%\cite{Ho:2019org}

\bibitem{Ho:2019org} J.~Ho, R.~Berg, T.~G.~Steele, W.~Chen and D.~Harnett,
%``Is the $Y(2175)$ a Strangeonium Hybrid Meson?,''
Phys.\ Rev.\ D \textbf{100}, 034012 (2019).
%doi:10.1103/PhysRevD.100.034012 [arXiv:1905.12779 [hep-ph]].
%%CITATION = doi:10.1103/PhysRevD.100.034012;

%\cite{Wang:2006ri}

\bibitem{Wang:2006ri} Z.~G.~Wang,
%``Analysis of the Y(2175) as a tetraquark state with QCD sum rules,''
Nucl.\ Phys.\ A \textbf{791}, 106 (2007).
%doi:10.1016/j.nuclphysa.2007.04.012 [hep-ph/0610171].
%%CITATION = doi:10.1016/j.nuclphysa.2007.04.012;%%
%91 citations counted in INSPIRE as of 24 May 2019

%\cite{Chen:2008ej}

\bibitem{Chen:2008ej} H.~X.~Chen, X.~Liu, A.~Hosaka and S.~L.~Zhu,
%``The Y(2175) State in the QCD Sum Rule,''
Phys.\ Rev.\ D \textbf{78}, 034012 (2008).
%doi:10.1103/PhysRevD.78.034012 [arXiv:0801.4603 [hep-ph]].
%%CITATION = doi:10.1103/PhysRevD.78.034012;%%
%35 citations counted in INSPIRE as of 24 May 2019

%\cite{Chen:2018kuu}

\bibitem{Chen:2018kuu} H.~X.~Chen, C.~P.~Shen and S.~L.~Zhu,
%``A possible partner state of the $Y(2175)$,''
Phys.\ Rev.\ D \textbf{98}, 014011 (2018).
%doi:10.1103/PhysRevD.98.014011 [arXiv:1805.06100 [hep-ph]].
%%CITATION = doi:10.1103/PhysRevD.98.014011;%%
%8 citations counted in INSPIRE as of 24 May 2019

%\cite{Cui:2019roq}

\bibitem{Cui:2019roq} E.~L.~Cui, H.~M.~Yang, H.~X.~Chen, W.~Chen and
C.~P.~Shen,
%``QCD sum rule studies of $s s {\bar{s}} {\bar{s}}$ tetraquark states with $J^{PC} = 1^{+-}$,''
Eur.\ Phys.\ J.\ C \textbf{79}, 232 (2019).
%doi:10.1140/epjc/s10052-019-6755-y [arXiv:1901.01724 [hep-ph]].
%%CITATION = doi:10.1140/epjc/s10052-019-6755-y;%%6 citations

%\cite{Wang:2019nln}

\bibitem{Wang:2019nln} Z.~G.~Wang, %``Light tetraquark state candidates,''
arXiv:1901.04815 [hep-ph]. %%CITATION = ARXIV:1901.04815;%%
%3 citations counted in INSPIRE as of 24 May 2019

%\cite{Azizi:2019ecm}

\bibitem{Azizi:2019ecm} K.~Azizi, S.~S.~Agaev, and H.~Sundu, Nucl.\ Phys.\ B
\textbf{948}, 114789 (2019).

%\cite{Jaffe:1976ig}

\bibitem{Jaffe:1976ig} R.~L.~Jaffe,
%``Multi-Quark Hadrons. 1. The Phenomenology of (2 Quark 2 anti-Quark) Mesons,''
Phys.\ Rev.\ D \textbf{15}, 267 (1977). %%doi:10.1103/PhysRevD.15.267
%%CITATION = doi:10.1103/PhysRevD.15.267;%%
%%1865 citations counted in INSPIRE as of 18 Oct 2017

%\cite{Kim:2017yvd}

\bibitem{Kim:2017yvd} H.~Kim, K.~S.~Kim, M.~K.~Cheoun and M.~Oka,
%``Tetraquark mixing framework for isoscalar resonances in light mesons,''
Phys.\ Rev.\ D \textbf{97}, 094005 (2018).
%%doi:10.1103/PhysRevD.97.094005  [arXiv:1711.08213 [hep-ph]].
%%CITATION = doi:10.1103/PhysRevD.97.094005;%%
%4 citations counted in INSPIRE as of 14 Jul 2018

%\cite{Agaev:2017cfz}

\bibitem{Agaev:2017cfz} S.~S.~Agaev, K.~Azizi and H.~Sundu,
%``The structure, mixing angle, mass and couplings of the light scalar $f_0(500)$ and $f_0(980)$ mesons,''
Phys.\ Lett.\ B \textbf{781}, 279 (2018).
%doi:10.1016/j.physletb.2018.03.085  [arXiv:1711.11553 [hep-ph]].  %%CITATION = doi:10.1016/j.physletb.2018.03.085;%%  %5 citations counted in INSPIRE as of 14 Jul 2018

%\cite{Agaev:2018sco}

\bibitem{Agaev:2018sco} S.~S.~Agaev, K.~Azizi and H.~Sundu,
%``The strong decays of the light scalar mesons $f_0(500)$ and $f_0(980)$,''
Phys.\ Lett.\ B \textbf{784}, 266 (2018).
%doi:10.1016/j.physletb.2018.07.042 [arXiv:1804.01726 [hep-ph]].
%%CITATION = doi:10.1016/j.physletb.2018.07.042;%%
%2 citations counted in INSPIRE as of 27 May 2019

%\cite{Shifman:1978bx}

\bibitem{Shifman:1978bx} M.~A.~Shifman, A.~I.~Vainshtein and V.~I.~Zakharov,
%``QCD and Resonance Physics. Theoretical Foundations,''
Nucl.\ Phys.\ B \textbf{147}, 385 (1979).
%doi:10.1016/0550-3213(79)90022-1  %%CITATION = doi:10.1016/0550-3213(79)90022-1;%%  %4985 citations counted in INSPIRE as of 11 Apr 2018

%\cite{Shifman:1978by}

\bibitem{Shifman:1978by} M.~A.~Shifman, A.~I.~Vainshtein and V.~I.~Zakharov,
%``QCD and Resonance Physics: Applications,''
Nucl.\ Phys.\ B \textbf{147}, 448 (1979).
%doi:10.1016/0550-3213(79)90023-3  %%CITATION = doi:10.1016/0550-3213(79)90023-3;%%  %2767 citations counted in INSPIRE as of 11 Apr 2018

%\cite{Balitsky:1989ry}

\bibitem{Balitsky:1989ry} I.~I.~Balitsky, V.~M.~Braun and
A.~V.~Kolesnichenko,
%``Radiative Decay Sigma+ ---> p gamma in Quantum Chromodynamics,''
Nucl.\ Phys.\ B \textbf{312}, 509 (1989). %doi:10.1016/0550-3213(89)90570-1
%%CITATION = doi:10.1016/0550-3213(89)90570-1;%%  %436 citations counted in INSPIRE as of 16 Sep 2017

%\cite{Belyaev:1994zk}

\bibitem{Belyaev:1994zk} V.~M.~Belyaev, V.~M.~Braun, A.~Khodjamirian and
R.~Ruckl, %``D* D pi and B* B pi couplings in QCD,''
Phys.\ Rev.\ D \textbf{51}, 6177 (1995).
%doi:10.1103/PhysRevD.51.6177  [hep-ph/9410280].  %%CITATION = doi:10.1103/PhysRevD.51.6177;%%  %412 citations counted in INSPIRE as of 17 Apr 2016

%\cite{Ioffe:1983ju}

\bibitem{Ioffe:1983ju} B.~L.~Ioffe and A.~V.~Smilga,
%``Nucleon Magnetic Moments and Magnetic Properties of Vacuum in QCD,''
Nucl.\ Phys.\ B \textbf{232}, 109 (1984).
%%CITATION = doi:10.1016/0550-3213(84)90364-X;%%

%\cite{Agaev:2016dev}

\bibitem{Agaev:2016dev} S.~S.~Agaev, K.~Azizi and H.~Sundu,
%Strong $Z_c^{+}(3900)\rightarrow J/\psi \pi^{+}; \eta_{c} \rho^{+}$ decays in QCD,
Phys.\ Rev.\ D\ \textbf{93}, 074002 (2016).

%\cite{Jaffe:2004ph}

\bibitem{Jaffe:2004ph} R.~L.~Jaffe, %``Exotica,''
Phys.\ Rept.\ \textbf{409}, 1 (2005).
% doi:10.1016/j.physrep.2004.11.005 [hep-ph/0409065].
%%CITATION = doi:10.1016/j.physrep.2004.11.005;%%
%359 citations counted in INSPIRE as of 21 Jun 2016

%\cite{Chen:2010ze}

\bibitem{Chen:2010ze} W.~Chen and S.~L.~Zhu,
%``The Vector and Axial-Vector Charmonium-like States,''
Phys.\ Rev.\ D \textbf{83}, 034010 (2011).
%doi:10.1103/PhysRevD.83.034010  [arXiv:1010.3397 [hep-ph]].  %%CITATION = doi:10.1103/PhysRevD.83.034010;%%  %81 citations counted in INSPIRE as of 09 Apr 2018

%\cite{Chen:2007xr}

\bibitem{Chen:2007xr} H.~X.~Chen, A.~Hosaka and S.~L.~Zhu,
%``Light Scalar Tetraquark Mesons in the QCD Sum Rule,''
Phys.\ Rev.\ D \textbf{76}, 094025 (2007). %doi:10.1103/PhysRevD.76.094025
%[arXiv:0707.4586 [hep-ph]].
%%CITATION = doi:10.1103/PhysRevD.76.094025;%%
%64 citations counted in INSPIRE as of 28 Oct 2017

%\cite{Maiani:2014}

\bibitem{Maiani:2014} L.~Maiani, F.~Piccinini, A.~D.~Polosa and V.~Riquer,
%``The Z(4430) and a new paradigm for spin interactions in tetraquarks,''
Phys.\ Rev.\ D \textbf{89}, 114010 (2014).

%\cite{Wang:2014vha}

\bibitem{Wang:2014vha} Z.~G.~Wang,
%``Analysis of the $Z(4430)$ as the first radial excitation of the $Z_c(3900)$,''
Commun.\ Theor.\ Phys.\ \textbf{63}, 325 (2015).
%doi:10.1088/0253-6102/63/3/325  [arXiv:1405.3581 [hep-ph]].  %%CITATION = doi:10.1088/0253-6102/63/3/325;%%  %25 citations counted in INSPIRE as of 01 Apr 2017

%\cite{Agaev:2017tzv}

\bibitem{Agaev:2017tzv} S.~S.~Agaev, K.~Azizi and H.~Sundu,
%``Treating $Z_c(3900)$ and $Z(4430)$ as the ground-state and first radially excited tetraquarks,''
Phys.\ Rev.\ D \textbf{96}, 034026 (2017).
%doi:10.1103/PhysRevD.96.034026  [arXiv:1706.01216 [hep-ph]].  %%CITATION = doi:10.1103/PhysRevD.96.034026;%%  %11 citations counted in INSPIRE as of 22 Sep 2018

%\cite{Ball:1996tb}

\bibitem{Ball:1996tb} P.~Ball and V.~M.~Braun,
%``The Rho meson light cone distribution amplitudes of leading twist revisited,''
Phys.\ Rev.\ D \textbf{54}, 2182 (1996).
%doi:10.1103/PhysRevD.54.2182 [hep-ph/9602323].
%%CITATION = doi:10.1103/PhysRevD.54.2182;%%

%\cite{Ball:1998sk}

\bibitem{Ball:1998sk} P.~Ball, V.~M.~Braun, Y.~Koike and K.~Tanaka,
%``Higher twist distribution amplitudes of vector mesons in QCD: Formalism and twist - three distributions,''
Nucl.\ Phys.\ B \textbf{529}, 323 (1998).
%doi:10.1016/S0550-3213(98)00356-3 [hep-ph/9802299].
%%CITATION = doi:10.1016/S0550-3213(98)00356-3;%%

%\cite{Ball:1998ff}

\bibitem{Ball:1998ff} P.~Ball and V.~M.~Braun,
%``Higher twist distribution amplitudes of vector mesons in QCD: Twist - 4 distributions and meson mass corrections,''
Nucl.\ Phys.\ B \textbf{543}, 201 (1999).
%doi:10.1016/S0550-3213(99)00014-0 [hep-ph/9810475].
%%CITATION = doi:10.1016/S0550-3213(99)00014-0;%%

%\cite{Ball:2007rt}

\bibitem{Ball:2007rt} P.~Ball and G.~W.~Jones,
%``Twist-3 distribution amplitudes of K* and phi mesons,''
JHEP \textbf{0703}, 069 (2007).
%doi:10.1088/1126-6708/2007/03/069 [hep-ph/0702100 [HEP-PH]].
%%CITATION = doi:10.1088/1126-6708/2007/03/069;%%

%\cite{Ball:2007zt}

\bibitem{Ball:2007zt} P.~Ball, V.~M.~Braun and A.~Lenz,
%``Twist-4 distribution amplitudes of the K* and phi mesons in QCD,''
JHEP \textbf{0708}, 090 (2007).
%doi:10.1088/1126-6708/2007/08/090 [arXiv:0707.1201 [hep-ph]].
%%CITATION = doi:10.1088/1126-6708/2007/08/090;%%

%\cite{Agaev:2016srl}

\bibitem{Agaev:2016srl} S.~S.~Agaev, K.~Azizi and H.~Sundu,
%``Application of the QCD light cone sum rule to tetraquarks: the strong vertices $X_bX_b\rho$ and $X_cX_c\rho$,''
Phys.\ Rev.\ D \textbf{93}, 114036 (2016).
%doi:10.1103/PhysRevD.93.114036  [arXiv:1605.02496 [hep-ph]].  %%CITATION = doi:10.1103/PhysRevD.93.114036;%%  %8 citations counted in INSPIRE as of 08 Nov 2017

%\cite{Agaev:2014wna}

\bibitem{Agaev:2014wna} S.~S.~Agaev, V.~M.~Braun, N.~Offen, F.~A.~Porkert
and A.~Sch\"{a}fer,
%``Transition form factors $\gamma^*\gamma\to\eta$ and $\gamma^*\gamma\to\eta'$ in QCD,''
Phys.\ Rev.\ D \textbf{90}, 074019 (2014).
%doi:10.1103/PhysRevD.90.074019 [arXiv:1409.4311 [hep-ph]].
%%CITATION = doi:10.1103/PhysRevD.90.074019;%%
%37 citations counted in INSPIRE as of 03 Jun 2019

%\cite{Agaev:2015faa}

\bibitem{Agaev:2015faa} S.~S.~Agaev, K.~Azizi and H.~Sundu,
%``Strong $D^{*}_sD_{s}\eta^{(\prime)}$ and $B^{*}_sB_{s}\eta^{(\prime)}$ vertices from QCD light-cone sum rules,''
Phys.\ Rev.\ D \textbf{92}, 116010 (2015).
%doi:10.1103/PhysRevD.92.116010 [arXiv:1509.08620 [hep-ph]].
%%CITATION = doi:10.1103/PhysRevD.92.116010;%%
%4 citations counted in INSPIRE as of 03 Jun 2019

%\cite{Agaev:2016dsg}

\bibitem{Agaev:2016dsg} S.~S.~Agaev, K.~Azizi and H.~Sundu,
%``Open charm-bottom scalar tetraquarks and their strong decays,''
Phys.\ Rev.\ D \textbf{95}, 034008 (2017).
%doi:10.1103/PhysRevD.95.034008[arXiv:1611.00293 [hep-ph]].
%%CITATION = doi:10.1103/PhysRevD.95.034008;%%
%15 citations counted in INSPIRE as of 03 Jun 2019

%\cite{Tanabashi:2018oca}

\bibitem{Tanabashi:2018oca} M.~Tanabashi \textit{et al.} [Particle Data
Group], Phys.\ Rev.\ D \textbf{98}, 030001 (2018).
\end{thebibliography}
\end{document}